\documentclass[a4paper,11pt]{article}

\usepackage{jheppub}
\usepackage{amsbsy}		% Bolds (vectors, tensors, etc.)
\usepackage{amsfonts}		% AMS fonts
\usepackage{amsmath}		% AMS maths
\usepackage{amssymb}		% AMS symbols
\usepackage{upgreek}		% upright Greek letters
\usepackage{wasysym}
\usepackage{multirow}

\usepackage{psfrag}
\usepackage{url}
\usepackage{float}

\newcommand{\tb}{\tan\beta}
\newcommand{\sinb}{\sin\beta}
\newcommand{\cosb}{\cos\beta}
\newcommand{\ctb}{\cot\beta}
\newcommand{\sina}{\sin\alpha}
\newcommand{\cosa}{\cos\alpha}
\newcommand{\sba}{\sin(\beta-\alpha)}
\newcommand{\cba}{\cos(\beta-\alpha)}

\newcommand{\gev}{~\mathrm{GeV}}
\newcommand{\tev}{~\mathrm{TeV}}
\newcommand{\ifb}{~\mathrm{fb}^{-1}}
\newcommand{\CP}{\mathcal{CP}}

\begin{document}
\hfill {\tt CERN-TH-2017-137}

\hfill {\tt SCIPP 17/07}

\def\thefootnote{\fnsymbol{footnote}}

\begin{center}
\Large\bf\boldmath
\vspace*{0.5cm}
 Status of the Charged Higgs Boson\\[0.4cm] in Two Higgs Doublet Models
\unboldmath
\end{center}
\vspace{0.4cm}
\begin{center}
A.~Arbey$^{1,2,3,}$\footnote{Also Institut Universitaire de France, 103 boulevard Saint-Michel, 75005 Paris, France}$^,$\footnote{Electronic address: alexandre.arbey@ens-lyon.fr}, F.~Mahmoudi$^{1,2,3,*,}$\footnote{Electronic address: nazila@cern.ch}, 
O.~St{\aa}l$^{4,}$\footnote{Electronic address: dr.oscar.stal@gmail.com}, T. Stefaniak$^{5,}$\footnote{Electronic address: tistefan@ucsc.edu}\\[0.4cm]
\vspace{0.4cm}
{\sl$^1$Univ Lyon, Univ Lyon 1, ENS de Lyon, CNRS, Centre de Recherche Astrophysique de Lyon UMR5574, F-69230 Saint-Genis-Laval, France}\\[0.2cm]
{\sl $^2$ Univ Lyon, Univ Lyon 1, CNRS/IN2P3, Institut de Physique Nucl\'eaire de Lyon UMR5822, F-69622 Villeurbanne, France}\\[0.2cm]
{\sl $^3$ CERN, Theoretical Physics Department, CH-1211 Geneva 23, Switzerland}\\[0.2cm]
{\sl $^4$ Lakegatan 6, SE-723 49 Westeros, Sweden}\\[0.2cm]
{\sl $^5$ Santa Cruz Institute For Particle Physics (SCIPP) and Department of Physics,\\ University of California, Santa Cruz, 1156 High Street, Santa Cruz, CA 95064, USA}\\[0.2cm]
\end{center}
\vspace{0.7cm}

\renewcommand{\thefootnote}{\arabic{footnote}}
\setcounter{footnote}{0}

\begin{abstract}

The existence of charged Higgs boson(s) is inevitable in models with two (or more) Higgs doublets. Hence, their discovery would constitute unambiguous evidence for new physics beyond the Standard Model (SM). 
Taking into account all relevant results from direct charged and neutral Higgs boson searches at LEP and the LHC, as well as the most recent constraints from flavour physics, we present a detailed analysis of the current phenomenological status of the charged Higgs sector in a variety of well-motivated Two Higgs Doublet Models (2HDMs).
We find that charged Higgs bosons as light as $75\gev$ can still be compatible with the combined data, although this implies severely suppressed charged Higgs couplings to all fermions. In more popular models, e.g.\ the 2HDM of Type II, we find that flavour physics observables impose a combined lower limit on the charged Higgs mass of $M_{H^\pm} \gtrsim 600$ GeV -- independent of $\tan\beta$ -- which increases to $M_{H^\pm} \gtrsim 650$ GeV for $\tan\beta < 1$. We furthermore find that in certain scenarios, the signature of a charged Higgs  boson decaying into a lighter neutral Higgs boson and a $W$ boson provides a promising experimental avenue that would greatly complement the existing LHC search programme for charged Higgs boson(s).

\end{abstract}
\newpage

%%%%%%%%%%%%%%%%%%%%%%%%%%%%%%%%%%%%%%%%%%%
\section{Introduction}

Following the discovery by ATLAS \cite{Aad:2012tfa} and CMS \cite{Chatrchyan:2012xdj} of a Higgs boson compatible with the predictions of the Standard Model (SM), the focus of experimental Higgs searches at the Large Hadron Collider (LHC) has shifted on the one hand towards performing precision Higgs rate measurements, in order to establish indirect evidence for physics beyond the SM (BSM), and on the other hand towards discovering other neutral or charged Higgs bosons, which are generically present in BSM theories with an extended Higgs sector. 

A well motivated class of models, compatible with the Higgs discovery, is given by extending the SM Higgs sector by a second scalar $\mathrm{SU}(2)_L$ doublet, the so-called Two Higgs Doublet Model (2HDM). The minimal supersymmetric extension of the SM (MSSM) is a prominent example of BSM theories that features a Higgs sector with two Higgs doublets. In particular, if all supersymmetric partners are heavy, the MSSM can be treated as an effective 2HDM for energies far below the SUSY mass scale(s)~\cite{Carena:1999py,Davidson:2005cw,Lee:2015uza,Bagnaschi:2015pwa}. Indeed, the 2HDM is one of the simplest low-energy effective Higgs models and could be the result of various UV-complete BSM theories, with or without Supersymmetry (SUSY).
The fact that no additional Higgs states have thus far been discovered constitutes a further motivation for thorough investigations of such scenarios. Phenomenologically, the 2HDM serves as a very useful and widely studied extension of the SM Higgs sector. It exhibits a plethora of collider signatures for LHC BSM Higgs searches and features interesting effects on low-energy (flavour) physics (see for example~\cite{Mahmoudi:2009zx,Maniatis:2009by,Jung:2010ik,Enberg:2011qh,Branco:2011iw,Arhrib:2011wc,Akeroyd:2012yg,Basso:2012st,Chang:2012ve,Chen:2013kt,Chiang:2013ixa,Crivellin:2013wna,Coleppa:2013dya,Chen:2013rba,Chang:2013ona,Li:2013vlx,Dumont:2014wha,Enberg:2014pua,Biswas:2014uba,Haber:2015pua,Bernon:2015qea,Bernon:2015wef,Keus:2015hva,Bernon:2015wef,Biswas:2015zgk,Bian:2016awe,Akeroyd:2016ymd,Cacciapaglia:2016tlr,Bhatia:2017ttp,Chakrabarty:2017qkh,Gori:2017qwg,Muhlleitner:2017dkd,Arhrib:2017wmo}), as we shall discuss in this work. 

Instead of considering the global picture after the $125$ GeV Higgs discovery, which has been the aim of many studies in the past few years (see for example \cite{Arbey:2011ab,Arbey:2011aa,Heinemeyer:2011aa,Bechtle:2012jw,Arbey:2012dq,Cabrera:2012vu,Altmannshofer:2012ks,Arbey:2012bp,Nath:2012nh,Carena:2013ytb,Arbey:2013jla,Liu:2013ula,Alanne:2013dra,Arbey:2015aca,Bhattacherjee:2015sga,Bechtle:2015nua,Bechtle:2016kui,Aggleton:2016tdd}), 
we focus in this work on one specific aspect of theories with additional Higgs doublets, namely, the presence of a charged Higgs boson, $H^\pm$. Charged scalar particles are a generic prediction of models where the SM Higgs sector is extended by at least one scalar $SU(2)_L$ doublet (or higher representations). Therefore, a charged Higgs discovery would be a clear signal of the existence of new physics beyond the Standard Model. It is therefore interesting to assess the current phenomenological constraints on the charged Higgs sector of the 2HDM --- the minimal model in which charged Higgs bosons appear.

We define several 2HDM scenarios in order to study different phenomenological aspects of the charged and neutral Higgs sectors: Three rather restricted scenarios that are well-motivated from the standpoint of both UV-complete theories (e.g.~the MSSM) as well as of complementary phenomenological constraints, such as electroweak precision observables (EWPOs) and Higgs signal rate measurements; and two rather general scenarios, where the assumptions of the previous scenarios are relaxed, and either the light or the heavy $\CP$-even Higgs boson is interpreted as the discovered Higgs state at $\sim 125\gev$. For each scenario, we consider all possible choices of $Z_2$ symmetric Yukawa structures that are compatible with natural absence of tree-level flavour changing neutral currents (FCNC).

We analyse the constraints posed by direct $H^\pm$ searches at LEP and the LHC, taking into account the most recent results (including the available results from Run 2 of the LHC). Furthermore, we investigate the implications of neutral Higgs boson searches for the charged Higgs sector. Clearly, these constraints on the charged Higgs sector depend on the assumed correlation of the parameters governing the neutral and charged Higgs sectors, and are therefore intrinsically model-dependent.

In addition to its distinct signatures in direct searches, the charged Higgs boson has a potentially very important role in flavour physics. Even in theories with minimal flavour violation (MFV) \cite{Chivukula:1987py,Hall:1990ac,DAmbrosio:2002vsn,Hurth:2012jn}, light $H^\pm$ exchange can give contributions to several observables at the same level as the SM predictions. The precise measurements of rare decay modes recorded at the level of the SM predictions therefore lead to lower limits on $M_{H^{\pm}}$. Based on the latest measurements of the branching fractions of $B\to X_s\gamma$, $B_s\to \mu^+\mu^-$, $B \to \tau\nu$, $B \to D^{(*)}\tau\nu$, and $\Delta M_{B_s}$ from LHC and $B$ factories, in addition to the recent LHCb results for the $B\to K^{(*)}\ell^+\ell^-$ and $B_s\to \phi\mu^+\mu^-$ decays in which some tensions with the SM predictions are observed, we perform an up-to-date analysis of the combined flavour constraints relevant for charged Higgs bosons in the general 2HDM, and emphasise the impact of the recent measurements.

This paper, which is a continuation and update of~\cite{Mahmoudi:2009zx}, is organised as follows. Section~\ref{sec:thdm} describes the theoretical framework of the two Higgs Doublet Model. In section~\ref{sec:experimental}, experimental constraints from LHC heavy Higgs searches and 125 GeV Higgs measurements as well as flavour observables are reviewed. Section~\ref{sec:2hdmscenarios} presents the different scenarios that are used in this analysis. Results are presented in section~\ref{sec:results} and conclusions are given in section~\ref{sec:conclusion}.

%%%%%%%%%%%%%%%%%%%%%%%%%%%%%%%%%%%%%%%%%%%
\section{The two Higgs doublet model (2HDM)}
\label{sec:thdm}

In the general 2HDM two identical Higgs doublets $\Phi_1$ and $\Phi_2$ with hypercharge $Y=+1/2$ are introduced. Assuming a softly-broken $Z_2$-symmetry, $\Phi_1 \to +\Phi_1$ and $\Phi_2 \to -\Phi_2$, on the possible quartic interactions, as well as $\CP$ conservation, the scalar potential can be written as
\begin{equation}
 \begin{aligned}
   V_{\mathrm{2HDM}}=    &\,m_{11}^2\Phi_1^\dagger\Phi_1+m_{22}^2\Phi_2^\dagger\Phi_2
    -\left[m_{12}^2\Phi_1^\dagger\Phi_2+\mathrm{h.c.}\right]
    +\tfrac{1}{2}\lambda_1(\Phi_1^\dagger \Phi_1)^2
    +\tfrac{1}{2}\lambda_2(\Phi_2^\dagger \Phi_2)^2\\
    &+\lambda_3(\Phi_1^\dagger \Phi_1)(\Phi_2^\dagger \Phi_2)
    +\lambda_4\left(\Phi_1^\dagger\Phi_2\right)\left(\Phi_2^\dagger\Phi_1\right)
    +\left\{
    \tfrac{1}{2}\lambda_5\left(\Phi_1^\dagger\Phi_2\right)^2
    +\mathrm{h.c.}\right\},
  \end{aligned}
\end{equation}
where all mass and coupling parameters can be chosen to be real. We impose that the minimum of the scalar potential preserves the $U(1)_\text{EM}$ gauge symmetry of electromagnetism, such that the scalar fields develop the following vacuum expectation values (vevs):
\begin{align}
\langle \Phi_1 \rangle = \frac{1}{\sqrt{2}} \left( \begin{array}{c} 0 \\ v_1 \end{array} \right), \qquad \langle \Phi_2 \rangle = \frac{1}{\sqrt{2}} \left( \begin{array}{c} 0 \\ v_2 \end{array} \right).
\end{align}
Using the minimisation conditions, the two mass parameters $m_{11}$ and $m_{22}$ can be traded for $v_1$ and $v_2$ or, more typically, $v=\sqrt{v_1^2+v_2^2}  \simeq 246\gev$ and $\tan\beta \equiv v_2/v_1$. The two Higgs doublets can be expanded around the potential minimum in terms of their component fields as
\begin{equation}
\Phi_1 = \left(
\begin{array}{c}
\phi_1^+\\
\frac{1}{\sqrt{2}}\left(v\,\cosb+\phi_1^0\right)
\end{array}
\right)
\qquad
\Phi_2 = \left(
\begin{array}{c}
\phi_2^+\\
\frac{1}{\sqrt{2}}\left(v\,\sinb+\phi_2^0\right)
\end{array}
\right).
\end{equation}

From the original eight scalar degrees of freedom, three Goldstone bosons ($G^\pm$ and $G$) are absorbed by the $W^\pm$ and $Z$ bosons. The remaining five degrees of freedom form the physical Higgs states of the model: two $\CP$-even scalars ($h$ and $H$, with masses $M_h$ and $M_H$, respectively, and $M_H \ge M_h$ per definition), one $\CP$-odd scalar ($A$), and a pair of charged Higgs bosons ($H^\pm$).

The squared-masses of the $\CP$-odd and charged Higgs states are
\begin{align}
M_A^2 &= \overline{m}^2 - \lambda_5 v^2\;,\\
M_{H^\pm}^2 &= M_A^2 + \tfrac{1}{2} v^2 (\lambda_5 - \lambda_4)\;, \label{eq:Mhp}
\end{align}
with $\overline{m}^2 \equiv 2 m_{12}^2/\sin(2\beta)$. 
The physical charged Higgs boson $H^+$ derives from the charged field component orthogonal to the direction of the vev, 
\begin{align}
H^+ = -\phi_1^+\sinb+\phi_2^+\cosb\;.
\end{align}
Similarly, the pseudoscalar Higgs boson is given by
\begin{align}
A = -\mathrm{Im}(\phi_1^0)\sinb+\mathrm{Im}(\phi_2^0)\cosb\;.
\end{align}

The two neutral $\CP$-even Higgs states mix, with the squared-mass matrix given by
\begin{align}
\arraycolsep=5pt
\def\arraystretch{1.3}
\mathcal{M}^2 = 
\begin{pmatrix}
\lambda_1 v^2 c_\beta^2 + \left(M_A^2 + \lambda_5v^2\right) s_\beta^2 & \left[\lambda_{345} v^2 - (M_A^2 + \lambda_5v^2)\right]s_\beta c_\beta \\
\left[\lambda_{345}v^2  - (M_A^2 + \lambda_5v^2)\right]s_\beta c_\beta &  \lambda_2 v^2 s_\beta^2 + \left(M_A^2 + \lambda_5v^2\right) c_\beta^2 
\label{eq:squaredmassmatrix}
\end{pmatrix},
\end{align}
where we defined $\lambda_{345} \equiv \lambda_3 + \lambda_4 + \lambda_5$.
For the diagonalisation of Eq.~\eqref{eq:squaredmassmatrix} we obtain the squared-masses of the $\CP$-even Higgs bosons,
\begin{align}
M_{H,h}^2 = \tfrac{1}{2} \left[ \mathcal{M}_{11}^2 +  \mathcal{M}_{22}^2 \pm \Delta \right],
\end{align}
with
\begin{align}
\Delta \equiv \sqrt{(\mathcal{M}_{11}^2 - \mathcal{M}_{22}^2)^2 + 4 (M_{12}^2)^2 }\, \ge \, 0\;,
\end{align}
and where the physical mass eigenstates are given by
\begin{align}
\arraycolsep=5pt
\def\arraystretch{1.3}
%This is correct:
\left(\begin{array}{c} h \\ H \end{array}\right) = 
\begin{pmatrix}
-  \sina & \cosa \\
\cosa & \sina
\end{pmatrix}
\left( \begin{array}{c}
\mathrm{Re}\, \phi_1^0\\
\mathrm{Re}\, \phi_2^0
\end{array}
\right).
\end{align}
The mixing angle $\alpha$ is defined modulo $\pi$. It is often useful to choose $|\alpha| \le \pi/2$, such that $\cosb \ge 0$. Then, assuming that $M_h\ne M_H$, the mixing angle is given by
\begin{align}
\cosa = \sqrt{\frac{\Delta + \mathcal{M}_{11}^2 - \mathcal{M}_{22}^2}{2\Delta}} = \sqrt{\frac{\mathcal{M}_{11}^2 - M_h^2}{M_H^2 - M_h^2}}\;.
\end{align}
It often proves convenient to express the scalar doublet fields in the \emph{Higgs basis}~\cite{Georgi:1978ri,Branco:1999fs,Davidson:2005cw}, defined by
\begin{align}
H_1 = \left(\begin{array}{c} H_1^+ \\ H_1^0 \end{array}\right) \equiv \Phi_1 \cosb + \Phi_2 \sinb\;, \qquad H_2 = \left(\begin{array}{c} H_2^+ \\ H_2^0 \end{array}\right) \equiv -\Phi_1 \sinb + \Phi_2 \cosb\;,
\label{eq:Higgsbasis}
\end{align}
such that the vacuum expectation values of these fields are $\langle H_1^0\rangle = v/\sqrt{2}$ and $\langle H_2^0 \rangle = 0$. Thus, the scalar doublet $H_1$ possesses the same tree-level couplings to all the SM particles as the SM Higgs boson. In the Higgs basis the physical Higgs states are given by
\begin{align}
\arraycolsep=5pt
\def\arraystretch{1.3}
\left(\begin{array}{c} h \\ H \end{array}\right) = 
\begin{pmatrix}
-  \sba & \cba \\
\cba & \sba
\end{pmatrix}
\left( \begin{array}{c}
\mathrm{Re}\, (H_1^0) - v\\
\mathrm{Re}\, (H_2^0)
\end{array}
\right).
\end{align}

If one of the physical Higgs states is \emph{aligned} with $\mathrm{Re}(H_1^0) - v$, it obtains the tree-level couplings of a SM Higgs boson. For the light Higgs state $h$ [heavy Higgs state $H$] this occurs when $\cos(\beta-\alpha)\to 0$ [$\sin(\beta-\alpha)\to 0$]. Thus, each case can provide a possible explanation of the $125\gev$ Higgs signal~\cite{Bernon:2015qea,Bernon:2015wef}. 

There is enough freedom in the model to choose all physical Higgs masses as input, which will be convenient for us in this work. In addition, we take the parameters $\sba$, $\tb$ and $m_{12}^2$ as input. For the latter parameter, which has no consequence on the charged Higgs boson sector, we shall typically assume the tree-level MSSM relation:
\begin{align}
m_{12}^2 = M_A^2 \tb/(1+\tan^2\beta)\;,
\label{eq:m12MSSM}
\end{align}
which is equivalent to $\lambda_5 = 0$ (see Sec.~\ref{sec:2hdmscenarios}).

The Higgs boson couplings to fermions are described by Yukawa interactions which in the general case are given by
\begin{equation}
-\mathcal{L}_{\mathrm{Yuk}} = \sum_{i=1}^2\left[\overline{Q}_L \widetilde{\Phi}_i\eta_i^U  U_R + \overline{Q}_L \Phi_i\eta_i^D  D_R + \overline{L}_L \Phi_i\eta_i^L  E_R+\mathrm{h.c.}\right],
\end{equation}
where $\widetilde{\Phi}=\mathrm{i}\sigma_2 \Phi$ and $\eta_i^F$ are $3\times 3$ Yukawa matrices in the flavour space for each Higgs doublet  ($i = 1,2$) and fermion class ($F=U,D,L$, for up-type quarks, down-type quarks and leptons, respectively). To obtain the combination corresponding to the mass matrix, we again perform a rotation by the angle $\beta$ which leads to new matrices
\begin{equation}
\begin{aligned}
\zeta^F_1 &= \eta^F_1 \cosb + \eta^F_2 \sinb\;,\\
\zeta^F_2 &= -\eta^F_1 \sinb + \eta^F_2 \cosb\;.
\end{aligned}
\label{eq:zeta}
\end{equation}
The new Yukawa matrices $\zeta^F_i$ ($i=1,2$) describe the couplings of the Higgs fields $H_i$ defined in the Higgs basis, Eq.~\eqref{eq:Higgsbasis}, to the SM fermions, i.e, the Higgs doublet field that couples with $\zeta^F_1$ acquires the SM Higgs vev $v$. After diagonalisation of the fermion mass eigenstates we can identify the mass matrices $M^F = \mathrm{diag}(\zeta_1^F)$.

If we restrict ourselves to the case with a $Z_2$ symmetry to prevent tree-level FCNCs, only one doublet couples to each class of fermions (i.e., either $\eta^F_1=0$ or $\eta^F_2=0$). We can then use Eq.~\eqref{eq:zeta} to define

\begin{equation}
\zeta^F_2 = \lambda^F \zeta^F_1.
\end{equation}
The assignment of $Z_2$ charges determines the pattern of the parameters $\lambda^F$. There are four physically distinct $Z_2$ charge assignments, which are labelled as 2HDM Types I--IV~\cite{Barger:1989fj}. In these cases, the $\lambda^F$ matrices are proportional to the unit matrix such that the diagonal matrix elements of $\lambda^F$ are equal to $\lambda_{FF}$. Our conventions for the types and the resulting $\lambda_{FF}$ are listed in Table~\ref{tab:Z2}.
\begin{table}
\centering
\begin{tabular*}{0.7\columnwidth}{@{\extracolsep{\fill}}ccccccc}
\hline
Type & $U_R$ & $D_R$ & $L_R$ & $\lambda_{UU}$ & $\lambda_{DD}$ & $\lambda_{LL}$ \\
\hline
I & $+$ & $+$ & $+$ & $\ctb$ & $\ctb$ & $\ctb$ \\
II & $+$ & $-$ & $-$ & $\ctb$ & $-\tb$ & $-\tb$ \\
III & $+$ & $-$ & $+$ & $\ctb$ & $-\tb$ & $\ctb$ \\
IV & $+$ & $+$ & $-$ & $\ctb$ & $\ctb$ & $-\tb$ \\
\hline
\end{tabular*}
\caption{Assignment of $Z_2$ charges for the right-handed fermions, and the resulting relations among Yukawa coupling matrices in the $Z_2$-symmetric types of 2HDM Yukawa sectors. The Higgs doublets $\Phi_1$ and $\Phi_2$ have $Z_2$ quantum numbers  $-$  and $+$, respectively.}
\label{tab:Z2}
\end{table}

The couplings of the charged Higgs boson to fermions, which are of relevance for our studies, are given by $-ig_{H^+\bar{f}f'}$, with
\begin{align}
g_{H^+\bar{u}_i d_j}&=\frac{V_{ij}^\text{CKM}}{\sqrt{2}M_W}\left[\lambda_{UU}\frac{1-\gamma_5}{2} + \lambda_{DD} \frac{1+\gamma_5}{2} \right],\\
%\end{equation}
%\begin{equation}
g_{H^+\bar{\nu}_\ell \ell}&=\frac{1}{\sqrt{2}M_W} \lambda_{LL} \frac{1+\gamma_5}{2}\;,
\end{align}
for the charged Higgs boson interactions with quarks and leptons, respectively. Here, $V^\text{CKM}$ is the Cabibbo-Kobayashi-Maskawa (CKM) matrix and $M_W$ is the $W$ boson mass.

It is also possible to relax the $Z_2$-symmetry, and consider generic Yukawa couplings, such that the $\lambda^F$ matrix elements are free parameters. In such a case, the charged Higgs couplings to fermions read:
\begin{align}
g_{H^+\bar{u}_i d_j}&=\frac{V_{ij}^\text{CKM}}{\sqrt{2}M_W}\left[\lambda^U_{ij}\frac{1-\gamma_5}{2} + \lambda^D_{ij} \frac{1+\gamma_5}{2} \right],\\
g_{H^+\bar{\nu}_\ell \ell'}&=\frac{1}{\sqrt{2}M_W} \lambda^L_{\ell\ell'} \frac{1+\gamma_5}{2}\;,
\end{align}
where the $\lambda^F_{ij}$ are free parameters. In the following, we will use the short-hand notations $\lambda_{d_id_j}\equiv\lambda^D_{ij}$, $\lambda_{u_iu_j}\equiv\lambda^U_{ij}$, $\lambda_{\ell\ell'}\equiv\lambda^L_{\ell\ell'}$. We also restrict our analysis to the case of flavour conserving couplings only where the $\lambda_{ij}$ vanish if $i\neq j$. We are then left with nine free parameters: $\lambda_{dd}$, $\lambda_{uu}$, $\lambda_{ss}$, $\lambda_{cc}$, $\lambda_{bb}$, $\lambda_{tt}$, $\lambda_{ee}$, $\lambda_{\mu\mu}$, $\lambda_{\tau\tau}$. 
Inspired by the hierarchy present in the $Z_2$-symmetric models, one can further restrict the $\lambda$'s to
\begin{equation}
\begin{aligned}
& m_d |\lambda_{dd}| \sim m_u|\lambda_{uu}|\sim 0\;,\\
& m_s |\lambda_{ss}|\sim m_c|\lambda_{cc}|\;,\qquad  m_b|\lambda_{bb}|\sim m_t|\lambda_{tt}|\;,\\
& m_e|\lambda_{ee}| \ll m_{\mu}|\lambda_{\mu\mu}| \ll m_{\tau}|\lambda_{\tau\tau}|\;.
\end{aligned}
\end{equation}
Nevertheless, in our analysis, we will consider the nine couplings to be free and constrain them with flavour observables.

%%%%%%%%%%%%%%%%%%%%%%%%%%%%%%%%%%%%%%%%%%%

\section{Experimental Constraints}
\label{sec:experimental}
\subsection{Collider limits}
\label{sec:colliderconstraints}
\subsubsection{Direct constraints from charged Higgs boson searches}
\label{sub:chargedcollider}
Direct constraints can be obtained from collider searches for the production and decay of on-shell charged Higgs bosons. These limits have the advantage of being very robust and model-independent, as long as the basic assumptions on the production and decay modes are satisfied. The search sensitivity is usually bounded by the kinematic reach of each experiment.

The LEP experiments have performed direct searches for charged Higgs bosons that are produced pairwise in Drell-Yan events, $e^+e^-\to \gamma/Z\to H^+H^-$. Since the coupling to photons is fixed by electromagnetic charge universality, the predictions for this process depend only on $M_{H^\pm}$. The resulting limit on the charged Higgs boson mass becomes robust when all relevant charged Higgs boson decay channels are considered. Combining data of the four LEP experiments from searches in the $\tau\nu$ and $cs$ final states, a limit of $M_{H^\pm} \gtrsim 80\gev$ (still mildly dependent on the branching ratio $\mathrm{BR}(H^\pm\to \tau\nu)$) is obtained under the assumption that the decay $H^\pm \to W^\pm h$, with a light neutral Higgs boson $h$, is absent~\cite{Abbiendi:2013hk}.\footnote{In fact, the experimental searches for charged Higgs bosons in the $cs$ final state are not exclusively sensitive to these particular quark flavours, i.e.\ the limit is also applicable for other light quark final states.} If the decay $H^\pm \to W^\pm h$ is open, dedicated LEP searches by the DELPHI~\cite{Abdallah:2003wd} and OPAL~\cite{Abbiendi:2008aa} collaborations for this signature (assuming the light Higgs having a mass $M_h \ge 12\gev$ and decaying to $b\bar{b}$) provide complementary constraints, such that the combined charged Higgs mass limit weakens only slightly, e.g.~$M_{H^\pm} \gtrsim 72.5$ GeV for $M_h=12$ GeV in Type-I models~\cite{Abbiendi:2013hk} (see also Ref.~\cite{Akeroyd:1998dt} for a phenomenological discussion of the $H^\pm\to W^\pm h$ decay in the context of LEP searches). In this work we include the combined LEP limits for the $\tau \nu$ and $cs$ final states~\cite{Abbiendi:2013hk}, as well as the OPAL limit for the $W^\pm h$ final state~\cite{Abbiendi:2008aa}, using \texttt{HiggsBounds} (version 5.1.0beta)~\cite{Bechtle:2008jh,Bechtle:2011sb,Bechtle:2013gu,Bechtle:2013wla,Bechtle:2015pma}.

At hadron colliders the charged Higgs boson searches can be categorised into two types, depending on the production mechanism. First, a light charged Higgs boson, with mass $M_{H^+}$, below the top quark mass, $m_t$, can be searched for in top quark decays, $t\to H^\pm b$; second, the charged Higgs boson can be produced directly in various production mechanisms, in particular, in association with a top and bottom quark, $pp\to H^\pm t b$. For a light charged Higgs boson with $M_{H^+} < m_t$ the first search type is typically more promising and has been in the focus of charged Higgs boson searches at the LHC during Run 1 with center-of-mass energies $\sqrt{s} = 7$ and $8\tev$, looking for charged Higgs bosons decaying to $\tau \nu$~\cite{Aad:2014kga,Khachatryan:2015qxa}, $c s$~\cite{Aad:2013hla,Khachatryan:2015uua} or $cb$~\cite{CMS:2016qoa}. With the increased center-of-mass energy $\sqrt{s}=13\tev$ of the LHC during Run 2, the direct charged Higgs production, $pp\to H^\pm t b$, becomes increasingly important, in particular, as it is sensitive to charged Higgs bosons heavier than the top quark. Thus far, the LHC experiments searched for $pp\to H^\pm t b$ in the charged Higgs boson decay modes $H^\pm \to \tau \nu$~\cite{Khachatryan:2015qxa,Aaboud:2016dig,ATLAS:2016grc,CMS:2016szv} and $H^\pm\to tb$~\cite{ATLAS:2016qiq} during Run 2, besides the aforementioned charged Higgs boson searches in top quark decays~\cite{CMS:2016szv}. Other charged Higgs boson searches have been performed in the vector boson fusion production channel with subsequent decay $H^\pm \to W^\pm Z$~\cite{Aad:2015nfa,Sirunyan:2017sbn}, however, due to the absence of the $H^\pm W^\mp Z$ coupling at tree-level in the 2HDM, these searches are irrelevant for our studies. 

The $95\%$~C.L.\ limits from all experimental searches mentioned above are included in our studies with \texttt{HiggsBounds}. We obtain the cross sections for the $pp\to H^\pm tb$ process at $8$ and $13\tev$ from grids provided by the LHC Higgs Cross Section Working Group (LHC~HXSWG)~\cite{deFlorian:2016spz}, which are based on Refs.~\cite{Berger:2003sm,Dittmaier:2009np,Flechl:2014wfa,Degrande:2015vpa,Degrande:2016hyf}. These grids can readily be used for 2HDMs with $Z_2$ symmetric Yukawa structures (Type I-IV) following a simple recipe for the $\tb$ rescaling. In the 2HDM scenario with a general Yukawa structure (scenario \textbf{(f)}, see Sec.~\ref{sec:2hdmscenarios}), we neglect the interference contribution $\propto \lambda_{tt} \lambda_{bb}$, and use only the contribution that is dominating, i.e. the cross section term $\propto \lambda_{tt}^2$ ($\propto  \lambda_{bb}^2$) if $m_t \lambda_{tt} >  m_b \lambda_{bb}$ ($m_t \lambda_{tt} <  m_b \lambda_{bb}$).

\subsubsection{Indirect constraints from neutral Higgs boson searches}

Collider searches for neutral Higgs bosons can also result in indirect, model-dependent limits on the charged Higgs boson. In theories were the Higgs masses and couplings are correlated, limits and measurements pertaining to the neutral Higgs bosons of the model can also constrain the properties of the charged Higgs boson. In this way, even parameter regions where direct $H^\pm$ production is insensitive or kinematically inaccessible can be constrained, albeit at the price of increased model-dependence. An example of a useful parameter correlation that relates the charged Higgs sector to the neutral Higgs sector in the general 2HDM is Eq.~\eqref{eq:Mhp}, which reduces to the well-known tree-level equality
\begin{align}
M_{H^\pm}^2=M_A^2+M_W^2 \label{eq:MSSMmasscorrelation}
\end{align}
in the MSSM. This relation can, in particular, result in very strong implications of pseudoscalar Higgs boson searches for the charged Higgs sector.

Searches for non-standard Higgs bosons are performed at the LHC in various channels with SM particles in the final state, i.e.~$\tau\tau$~\cite{ATLAS:2016fpj,CMS:2016rjp}, $b\bar{b}$~\cite{Malone:2015mia,Khachatryan:2015tra,CMS:2016ncz}, $t\bar{t}$~\cite{ATLAS:2016pyq}, $\mu\mu$~\cite{ATLAS:2016zzs,CMS:2016rme}, $\gamma\gamma$~\cite{ATLAS:2016eeo}, $WW$~\cite{Aaboud:2016okv,ATLAS:2016kjy,ATLAS:2016cwq,CMS:2016jpd}, $ZZ$~\cite{Aaboud:2016okv,ATLAS:2016oum,ATLAS:2016npe,CMS:2016ilx,CMS:2017sbi}, as well as in channels  with the SM-like Higgs boson  (denoted $h_\text{SM}$ here) in the final state, i.e.~$h_\text{SM} h_\text{SM}$~\cite{ATLAS:2016ixk,ATLAS:2016qmt,TheATLAScollaboration:2016ibb,Khachatryan:2015tha,CMS:2016tlj,CMS:2016vpz,CMS:2017orf,CMS:2017ums} and $h_\text{SM} Z$~\cite{Aad:2015wra,Khachatryan:2015tha,Khachatryan:2015lba,TheATLAScollaboration:2016loc}, with various decay modes of $h_\text{SM}$. CMS has also searched for processes involving two non-standard Higgs bosons, namely $H\to AZ$ or $A\to HZ$ decays (depending on the mass ordering)~\cite{Khachatryan:2016are,CMS:2016qxc}. These searches are specifically designed for probing 2HDM scenarios with large mass splittings between the Higgs bosons $H$ and $A$, as motivated from the standpoint of obtaining a strong first-order phase transition needed for successful electroweak baryogenesis~       \cite{Dorsch:2013wja,Dorsch:2014qja,Blinov:2015vma,Dorsch:2016tab,Basler:2016obg,Basler:2017uxn}. Lastly, there are searches for the SM-like Higgs boson decaying into lighter Higgs states, $h_\text{SM} \to h h$, with a light Higgs boson $h$ with a mass below $\sim 62.5\gev$ and decaying into $\mu\mu$, $\tau\tau$ or $bb$~\cite{Aaboud:2016oyb,Khachatryan:2017mnf,Khachatryan:2015wka,Khachatryan:2015nba,CMS:2016tgd}. 

All the above mentioned experimental constraints, with the notable exception of the non-standard Higgs boson search in the $t\bar{t}$ final state~\cite{ATLAS:2016pyq} due to the model-dependence of its limit, are included in our analysis via \texttt{HiggsBounds} at the $95\%$~C.L.

\subsubsection{Indirect constraints from measurements of the $125\gev$ Higgs boson}

Another type of indirect constraint on the charged Higgs boson comes from the mass and signal rate measurements of the Higgs boson discovered by the LHC experiments in 2012~\cite{Aad:2012tfa,Chatrchyan:2012xdj}, implied by the parameter correlations intrinsic to the model. A well-known example of such parameter correlations is again given by the tree-level MSSM Higgs sector, which is a 2HDM of Type~II with additional parameter constraints imposed by supersymmetry. Here, the coupling properties of the light Higgs boson become SM-like in the \emph{decoupling limit}, where the remaining Higgs states (including the charged Higgs boson due to Eq.~\eqref{eq:MSSMmasscorrelation}) are heavy with squared masses $ m ^2\gg v^2 \approx (246\gev)^2$~\cite{Gunion:2002zf}.\footnote{Note, however, that SM-like coupling properties of one of the $\CP$-even Higgs states can also be obtained in the \emph{alignment without decoupling} limit through an accidental cancellation of tree-level and loop-level effects~\cite{Carena:2013ooa,Dev:2014yca,Carena:2014nza,Bechtle:2016kui}. In this rather constrained scenario the remaining Higgs states can still be relatively light~\cite{Bechtle:2016kui,Profumo:2016zxo}.}

The combination of ATLAS and CMS data from Run 1~\cite{Khachatryan:2016vau} allowed to determine the Higgs mass to be $M_{h_\text{SM}} = 125.09 \pm 0.21 (\text{stat.}) \pm 0.11 (\text{syst.})\gev$, as well as to confirm  at the $\sim 10\%$ level that the discovered scalar exhibits the same signal rates as predicted by the SM.
Meanwhile, these results have been confirmed by ATLAS and CMS with the first $13\tev$ results~\cite{ATLAS:2016nke,ATLAS:2016oum,ATLAS:2016oum,ATLAS:2016gld,ATLAS:2016pkl,ATLAS:2016awy,ATLAS:2016ldo,CMS:2016ixj,CMS:2017jkd}, based on an integrated luminosity of up to $13.3\ifb$ and $12.9\ifb$, respectively. 

We include both the measured signal rates from the ATLAS and CMS Run 1 combination and the first $13\tev$ results in our study via the code \texttt{HiggsSignals-2.1.0beta}~\cite{Bechtle:2013xfa,Bechtle:2014ewa}.\footnote{A few results have already been shown with a larger integrated luminosity $\sim36\ifb$ per experiment~\cite{ATLAS:2017bkl,CMS:2017cbv,CMS:2017wyg,CMS:2017rli}, which we did not include in this study. These updated results again confirm the overall picture, but improve the precision of the signal rate determination only marginally. Therefore, including these updated measurements in our analysis would not lead to significant changes in our findings.} The code \texttt{HiggsSignals}  returns a $\chi^2$ value for the consistency between the model-predicted signal rates and the corresponding measurements, taking into account various correlations of systematic experimental and theoretical uncertainties (if known). Regarding the theoretical uncertainties, we follow the latest recommendations from the LHC HXSWG~\cite{deFlorian:2016spz} for the production cross section and decay width uncertainties to construct the relevant covariance matrix~\cite{Arbey:2016kqi}. We then determine the minimal $\chi^2$ value over the scanned parameter space, $\chi^2_\text{min}$, and regard as allowed the parameter space that features a $\chi^2$ value within $\Delta \chi^2 \equiv \chi^2 - \chi^2_\text{min} \le 5.99$ (which corresponds to a $95\%$ C.L.~in a two-dimensional (projected) parameter space in the Gaussian limit).

\subsection{Flavour constraints}
\label{sec:flav-constraints}
Flavour physics can impose stringent indirect constraints on new physics models, far beyond the mass scales reachable in direct searches. The constraining power for the properties of charged Higgs bosons is exceptionally good in some cases, e.g.\ with the $B_u \to \tau \nu$ decay, where the $H^\pm$ contribution enters already at tree-level \cite{Kalinowski:1990ba}.

\begin{table}[t]
\begin{center}
\begin{tabular}{|l|l|l|}\hline
Observable & Experiment & SM prediction \\ \hline
BR($B \to X_s \gamma$) & $(3.32 \pm 0.15)\times 10^{-4}$ \cite{Amhis:2016xyh} & $(3.34 \pm 0.22)\times 10^{-4}$\\ \hline
$\Delta_0(B \to K^* \gamma)$ & $(1.2 \pm 5.1)\times 10^{-2}$ \cite{Amhis:2016xyh}  & $(5.33 \pm 2.6)\times 10^{-2}$\\ \hline
BR($B_s \to \mu^+\mu^-$) & $ (3.0 \pm 0.6 \pm 0.25) \times 10^{-9}$ \cite{Aaij:2017vad} & $(3.54 \pm 0.27) \times 10^{-9}$\\ \hline
BR($B_u \to \tau\nu$) & $(1.06\pm 0.19)\times 10^{-4}$ \cite{Amhis:2016xyh} & $(0.82\pm 0.29)\times 10^{-4}$\\ \hline
BR($D_s\to \tau \nu$) & $(5.51 \pm 0.24)\times 10^{-2}$ \cite{Amhis:2016xyh} & $(5.13\pm 0.11)\times 10^{-2}$\\ \hline
% R$_{\mu23}$($K\to \mu \nu_\mu$) & 1 & \\ \hline
$\Delta M_{B_s}$ & $17.76 \pm 0.021$ ps$^{-1}$ \cite{Amhis:2016xyh} & $17.38 \pm 1.505$ ps$^{-1}$\\ \hline
\end{tabular}
\caption{Experimental results and the corresponding SM values for a selected number of flavour physics observables. The experimental data represents the most recent measurements or official combinations. %The SM predictions are evaluated with {\tt SuperIso}\ \cite{Mahmoudi:2007vz,*Mahmoudi:2008tp,*Mahmoudi:2009zz}.
The experimental values and SM predictions for the observables related to $B\to K^{(*)}\ell^+\ell^-$ and $B_s\to \phi\mu^+\mu^-$ can be found in Ref.~\cite{Hurth:2016fbr}.
}
\label{Inputobservables}
\end{center}
\end{table}

For the present analysis we have chosen a set of ``conventional'' flavour physics observables sensitive to charged Higgs exchange. These are listed in Table~\ref{Inputobservables} where we give the experimentally measured values and the corresponding SM predictions. 
In addition, we consider the recent measurements in the $B\to K^{(*)}\ell^+\ell^-$ and $B_s\to \phi\mu^+\mu^-$ decays, where LHCb has measured a multitude of clean angular observables in addition to the branching ratios. These measurements are performed in bins of the dilepton invariant mass, $q^2$.
Most of these observables are in good agreement with the SM predictions. However, some deviations from the SM predictions were observed in certain $q^2$ bins for some of the observables. The largest deviation with $\sim 3\sigma$ significance was reported in the $q^2 \in [4.0,6.0]$ and $[6.0,8.0]$ GeV$^2$ bins for the angular observable $P_5^\prime$~\cite{Aaij:2015oid}.\footnote{Recently, the Belle collaboration~\cite{Abdesselam:2016llu} as well as the ATLAS collaboration~\cite{ATLAS:2017dlm} also supported the deviation in $P_5^\prime$ although with larger experimental uncertainties, while the CMS collaboration reported a result in agreement with the SM prediction~\cite{CMS:2017ivg}.} Moreover, the branching ratio of the decay $B_s\to \phi \mu^+ \mu^-$ is found to be more than 3$\sigma$ below the SM predictions~\cite{Aaij:2015esa}. 
LHCb has in addition performed tests of lepton flavour universality by measuring the ratios 
$R_K = {\rm BR}(B^+ \to K^+ \mu^+ \mu^-)\,/\,{\rm BR}(B^+ \to K^+ e^+ e^-)$ 
and $R_{K^*} = {\rm BR}(B \to K^{*0} \mu^{+} \mu^{-})\,/\,{\rm BR}(B \to K^{*0} e^{+} e^{-})$
in the low-$q^2$ region
showing 2.6$\sigma$ and $\sim 2.5\sigma$ deviations respectively~\cite{Aaij:2014ora,Aaij:2017vbb}.
In this study, we include in a consistent way all these observables by performing a statistical combination taking into account both the theoretical and experimental correlations. The methodology and the list of observables with their definitions follow closely Refs.~\cite{Hurth:2016fbr,Chobanova:2017ghn,Hurth:2017hxg}.

\begin{table}[t]
\begin{center}
\begin{tabular}{|l|l|l|}\hline
Observable & Experiment & SM prediction \\ \hline
$\displaystyle\frac{\mathrm{BR}(B \to D \tau\nu)}{\mathrm{BR}(B \to D \ell\nu)}$ & $0.403 \pm 0.040 \pm 0.024$ \cite{Amhis:2016xyh} & $0.300 \pm 0.012$\\ \hline
$\displaystyle\frac{\mathrm{BR}(B \to D^* \tau\nu)}{\mathrm{BR}(B \to D^* \ell\nu)}$ & $0.310 \pm 0.015 \pm 0.08$ \cite{Amhis:2016xyh}  & $0.248 \pm 0.008$\\ \hline
\end{tabular}
\caption{$B \to D^{(*)}$ ratios used in this work. We consider -23\% experimental correlations between the two observables \cite{Amhis:2016xyh}.}
\label{ObsBDlnu}
\end{center}
\end{table}

Other deviations from the SM predictions have been reported by the Belle and BaBar experiments in the measurements of the ratios ${\rm BR}(B \to D \tau \nu)/{\rm BR}(B \to D \ell \nu)$ and ${\rm BR}(B \to D^* \tau \nu)/{\rm BR}(B \to D^* \ell \nu)$ (where $\ell$ is $e$ or $\mu$) that exceed the SM expectations by $\sim 4\sigma$~\cite{Amhis:2016xyh}. As the hadronic form factors tend to cancel in these ratios, they are rather sensitive to new physics contributions. A consistent explanation of the two ratios is not possible neither in the 2HDM Type~II nor in the MSSM, but it is for example possible in a non-standard 2HDM Type~III with generic couplings \cite{Crivellin:2012ye}. On the other hand, the recent Belle results~\cite{Abdesselam:2016xqt} are in agreement with the SM predictions within $0.6\sigma$. Therefore, while we shall comment on the impact of these measurements in the following, we do not include them in the global fits of the flavour observables. For the study of the $B \to D^{(*)} \ell\nu$ constraints, we use the values given in Table~\ref{ObsBDlnu} and consider an experimental correlation of $-23\%$ between the two observables \cite{Amhis:2016xyh}.

For the numerical evaluation of the SM predictions and calculation of the associated $\chi^2$, we use {\tt SuperIso v3.7} \cite{Mahmoudi:2007vz,*Mahmoudi:2008tp,*Mahmoudi:2009zz}, which is also used for the evaluation of the contributions beyond the SM in all models we analyse below. Ref.~\cite{Mahmoudi:2007vz,*Mahmoudi:2008tp,*Mahmoudi:2009zz} also contains an introduction to these observables, and the precise definitions that we employ.

%%%%%%%%%%%%%%%%%%%%%%%%%%%%%%%%%%%%%%%%%%%

\section{2HDM scenarios}
\label{sec:2hdmscenarios}

In this section we introduce the various 2HDM scenarios that are investigated in this work. As discussed in Sec.~\ref{sec:thdm},
in the 2HDM type I, all fermions couple to the same Higgs doublet (where all non-standard Higgs bosons, including the charged Higgs boson, decouple from the fermions for $\tb \gg 1$), type II gives the same $H^\pm$ couplings as in the MSSM at tree-level, type III has one doublet coupling to up-type quarks and leptons, while the other couples to down-type quarks, and finally in type IV one doublet couples to quarks and the other to leptons.

The charged Higgs phenomenology depends strongly on the type of Yukawa sector. For example, the coupling to third generation quarks is given by $g_{H^\pm \bar{t} b}\,\sim\, \left(m_b \,+\,  m_t\right)\cot\beta$ in the type I/III models, and $g_{H^\pm \bar{t} b}\,\sim\, m_b\tan\beta\, P_R \,+\, m_t\cot\beta\, P_L$ in the 2HDM type II/IV. 
Since the couplings to leptons (including the $\tau$) are also different for the four 2HDM types, different constraints on the charged Higgs mass and $\tan\beta$ are obtained from direct searches. Besides the changes in the $t\to bH^\pm$ decay rate, and the following charged Higgs decay modes, we also would like to point out that the differences in the vertex structure could potentially be probed by measuring the angular distributions of the $W^\pm/H^\pm$ decay products \cite{Eriksson:2007fx}. 

As discussed in Section~\ref{sec:thdm}, for each 2HDM type there are seven free parameters, which are in the physical basis
\begin{itemize}
 \item $M_h,M_H,M_A,M_{H^+}$, masses of the Higgs states,
 \item $\tb$, ratio of the Higgs doublet vevs,
 \item $\sba$, where $\alpha$ is the mixing angle of the $\CP$-even Higgs states,
 \item $m_{12}$, diagonal term of the mass matrix of the Higgs doublets.
\end{itemize}

\noindent In the following, we will consider six different illustrative scenarios and analyse the constraints from Higgs searches and flavour physics.

\begin{description}
 \item[(a)] \emph{MSSM-like regime}: We consider two free parameters, $M_{H^+}$ and $\tb$, and impose
 \begin{align*}
 M_H=M_A=\mathrm{max}(M_h,M_{H^+})\;,\quad  M_h=125.09\gev\;,\quad \sba=1\;,
% m_{12}^2=M_A^2 \tb/(1+\tan^2\beta).
 \end{align*}
 as well as Eq.~\eqref{eq:m12MSSM} to fix $m_{12}^2$. The constraint $\sba=1$ implies that the light Higgs boson $h$ has exactly the same tree-level couplings as the SM Higgs boson. By imposing Eq.~\eqref{eq:m12MSSM} and $M_H=M_A=M_{H^+}$, the 2HDM Type II is MSSM-like, i.e.\ its Higgs sector is similar to the tree-level MSSM Higgs sector in the decoupling limit, $M_A \gg M_Z$. The constraints in this scenario are therefore similar to the constraints on the MSSM Higgs sector provided the SUSY  particles are heavy.
 
 \item[(b)] \emph{Heavy neutral Higgs bosons}: In this scenario, we consider two free parameters, $M_{H^+}$ and $\tb$, and impose
 \begin{align*}
 M_H=M_A=1\tev\;,\quad  M_h=125.09\gev\;,\quad \sba=1\;,
 \end{align*}
 and Eq.~\eqref{eq:m12MSSM}. The only difference with scenario \textbf{(a)} is the fixed mass of the heavy neutral Higgs bosons $H$ and $A$, which allows us to relax the constraints from neutral Higgs searches and thus leads to a better understanding of the dependence of the charged Higgs and flavour constraints on the parameters.
 
 \item[(c)] \emph{Decoupling regime}: We allow for three free parameters, $M_{H^+}$, $M_H$ and $\tb$, and impose
 \begin{align*}
 M_H > \max(M_h, M_{H^+})\;, \quad M_h=125.09\gev\;, \quad M_A=M_{H^+}\;, \\
\cos(\beta-\alpha) = 0.1 \times (150\gev/M_H)^2\;,
 \end{align*}
 as well as Eq.~\eqref{eq:m12MSSM}. The choice of $\cos(\beta-\alpha)$ is made in order to approximately emulate the decoupling limit~\cite{Haber:2015pua}. 
 
 \item[(d)] \emph{General scenario}: We only set the light Higgs mass to $M_h=125.09\gev$ and vary the six remaining parameters, $M_H^+$, $M_H$, $M_A$, $\sba$, $\tb$ and $m_{12}^2$, imposing $M_H>M_h$. This is the most general scenario based on the assumption that the light Higgs state is the observed Higgs state. Studying the constraints in this scenario will allow us to obtain very general and conservative constraints from the Higgs searches and flavour physics on the charged Higgs sector.
 
 \item[(e)] \emph{Inverted scenario}: We consider the four parameters $M_{H^+}$, $M_h$, $\tb$ and $m_{12}^2$ to vary, and assume 
\begin{align*}
M_H=125.09\gev\;,\quad \sin(\beta-\alpha)=0\;,\quad M_A=M_{H^+}\;.%,\quad  M_h < M_H.
\end{align*}
Thus, the heavy $\CP$-even Higgs boson $H$ is assumed to be the observed Higgs state, so that this scenario is complementary to scenario \textbf{(d)}. The scan ranges for $M_h$ and $m_{12}^2$ are $M_h \in [50\gev, M_H]$
and $ m_{12}^2 \in [-2000^2, 2000^2]\gev^2$. For simplicity, we assume the $\CP$-odd Higgs boson $A$ to be mass degenerate with the charged Higgs boson $H^\pm$.

 \item[(f)] \emph{Generic couplings}: We set the light Higgs mass to $M_h=125.09\gev$ and vary the six remaining parameters, $M_H^+$, $M_H$, $M_A$, $\sba$, $\tb$ and $m_{12}^2$ (as in scenario \textbf{(d)}). In addition, the 9 flavour-conserving $\lambda_{ii}$ charged Higgs couplings are varied independently\footnote{For the $B\to D^{(*)}\ell\nu$ transitions, $\lambda_{cc}$, $\lambda_{bb}$ and $\lambda_{\tau\tau}$ are the only relevant generic couplings.} and are not related to $\tb$. This scenario is the most general $\CP$-conserving 2HDM scenario with minimal flavour violation.
 \end{description}

\noindent Scenarios \textbf{(a)}-\textbf{(c)} are specific cases, and the \textbf{(d)}-\textbf{(e)} are generic enough to allow for a test of the stability of the constraints. We will only consider scenario \textbf{(f)} in the context of the $B\to D^ {(*)}\ell\nu$ constraints.

%%%%%%%%%%%%%%%%%%%%%%%%%%%%%%%%%%%%%%%%%%%
\section{Results}
\label{sec:results}

\subsection{Higgs searches}
\label{sec:lhc}

In this section we  discuss the constraints from direct searches for charged Higgs bosons at the LHC, as well as the indirect constraints on the charged Higgs sector that can be inferred from LHC searches for neutral Higgs bosons, given certain model assumptions.

Indirect limits from neutral Higgs collider searches have a rather limited impact on the charged Higgs sector of the general 2HDM, because there is a large freedom in choosing several (or all) of the Higgs masses as free model parameters. Even if not all parameter choices lead to stable and perturbative configurations, it is generally not difficult to e.g.~achieve $M_h\sim 125\gev$ and $\sba=1$ without affecting $M_{H^\pm}$, thus ensuring very good agreement with the mass and rate measurements of the observed Higgs state. However, a light charged Higgs boson can affect the Higgs rates of loop-induced processes such as the decay $h\to \gamma\gamma$, as will be discussed in Sec.~\ref{sec:Higgsrates}. As another example, a large mass splitting within the second Higgs doublet can introduce large corrections to the oblique parameters~\cite{Peskin:1990zt,Peskin:1991sw} independently of the 2HDM Yukawa structure. In order to avoid this, an approximate custodial symmetry is desirable. This requirement either restricts $M_{H^\pm}$ to low values, $M_{H^\pm} \lesssim v$, or enforces one of the approximate relations $M_{H^\pm} \approx M_A$ or $M_{H^\pm}\approx M_H$. In the latter case, collider searches for non-standard Higgs bosons $H$ or $A$, including the MSSM searches for $H/A\to \tau\tau$, can then be used to infer constraints on the charged Higgs mass and other model parameters relevant to the charged Higgs sector (in particular, $\tb$). 

In the following we discuss the collider constraints for the scenarios \textbf{(a)}-\textbf{(e)} introduced in Sec.~\ref{sec:2hdmscenarios}.

\subsubsection{Specific scenarios (a)-(c)}
\label{sec:results_a-c}

We show the Higgs search constraints for scenario \textbf{(a)} (\emph{MSSM-like regime}) in Fig.~\ref{2hdm_mAmHmHp_HB} for the four Yukawa types. Recall that in this scenario, $M_{H^+} (=M_A=M_H)$ and $\tb$ are the only free parameters. The combined charged Higgs searches at LEP pose a robust lower charged Higgs mass limit of $M_{H^+} \ge 80\gev$, irrespective of the Yukawa type and $\tb$.\footnote{A light charged Higgs boson with $M_{H^+} \sim M_W \sim 80\gev$ can be motivated by a $2.8\sigma$ deviation from lepton universality found at LEP~\cite{Park:2006gk,Mader:2012pm}.} In fact, this $M_{H^+}$ limit becomes slightly stronger at large $\tb \gtrsim 5$ ($\tb$ between $5$ and $10$) for the Type I, III and IV (Type~II) models. The most relevant constraints from charged Higgs searches at the LHC are obtained from the processes $t\to H^\pm b$ with $H^\pm\to\tau\nu$ in the low $M_{H^+}$ regime ($M_{H^+} < m_t$), and $pp\to H^\pm tb$ with $H^\pm \to tb$ in the high $M_{H^+}$ regime ($M_{H^+} > m_t$). The former process is particularly relevant in Type~II, where charged Higgs masses below $\sim (155-160)\gev$ are quite robustly excluded (only mildly dependent on $\tan\beta$). For the latter process the experimental limit has only been presented for $M_{H^+} \ge 300\gev$, hence the sharp edge in the corresponding exclusion at $M_{H^+} = 300\gev$ in Fig.~\ref{2hdm_mAmHmHp_HB}.

 \begin{figure}[t!]
\centering
\includegraphics[width=0.48\columnwidth]{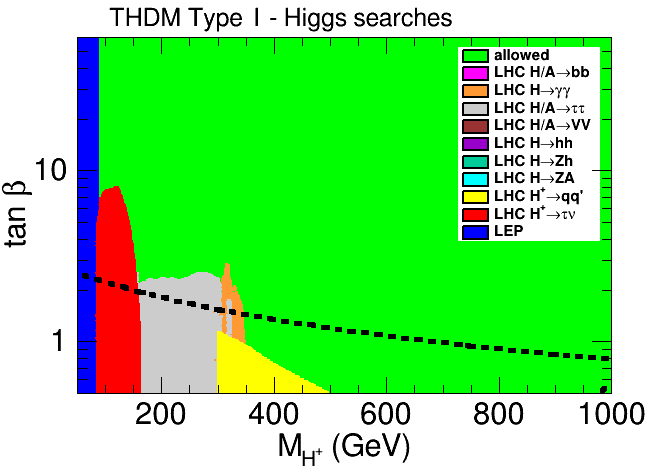}\hfill
\includegraphics[width=0.48\columnwidth]{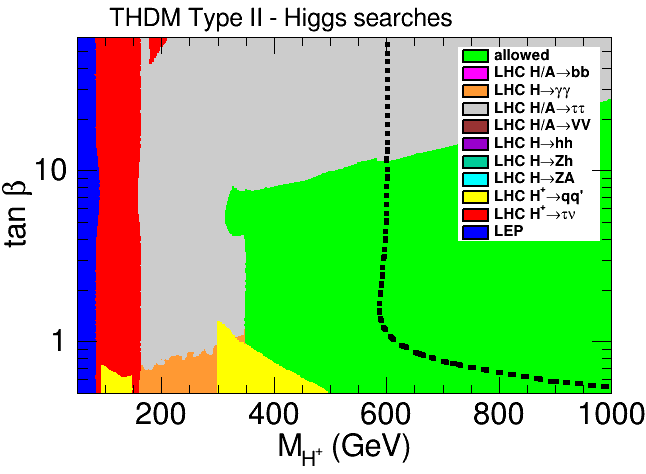}\\
\includegraphics[width=0.48\columnwidth]{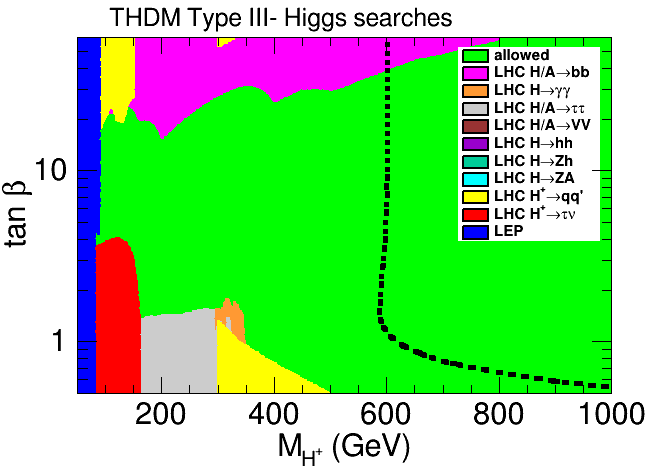}\hfill
\includegraphics[width=0.48\columnwidth]{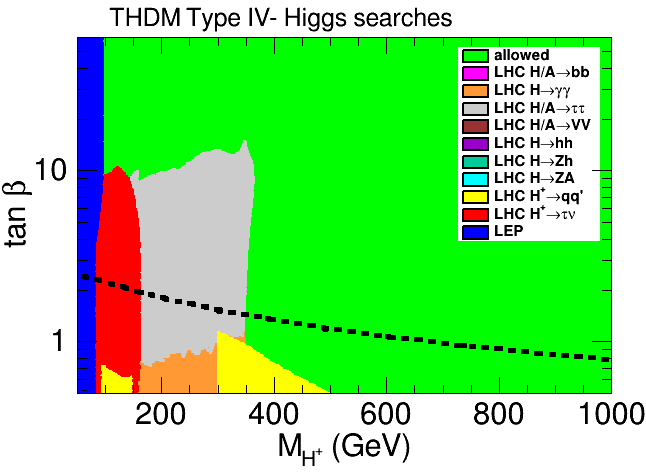}
\caption{Regions of the $(M_{H^+}, \tan\beta)$ parameter space of scenario \textbf{(a)} (\emph{MSSM-like scenario}) excluded at $95\%$ C.L.\ by charged and neutral Higgs searches (see Sec.~\ref{sec:colliderconstraints}) for the four different 2HDM Yukawa types. The colour coding corresponds to exclusion by different constraints, as given by the legend. The green region is consistent with all collider constraints. The dotted line shows the combined limit from all $b\to s$ observables (see Sec.~\ref{sub:flavour} for details).}
\label{2hdm_mAmHmHp_HB}
\end{figure}

Relevant constraints from neutral Higgs searches arise mostly from the processes $pp\to H/A\to\tau\tau$, and from $pp\to H\to\gamma\gamma$ at small $\tb$. The process $pp\to H/A\to bb$ is also important for Type~III at large $\tb$, because the $H/A$ couplings to bottom quarks are enhanced while the couplings to $\tau$-leptons are not. In Type~II, the LHC searches for $pp\to H/A\to \tau\tau$ impose strong constraints at large $\tb$, because the $H/A$ couplings to both bottom quarks and $\tau$-leptons are $\tb$ enhanced. At large $\tb$, the Higgs bosons are thus dominantly produced in association with bottom quarks in both Type~II and III, whereas the branching fraction for the $H/A\to \tau\tau$ decay is suppressed by the enhanced and dominant $H/A\to b\bar{b}$ decay only in Type~III, and it is not suppressed in Type~II. The  $pp\to H\to\gamma\gamma$ process can constrain the parameter space at low $\tb$ and $M_H < 2 m_t \approx 350\gev$, as the $\cot\beta$-enhanced $H$ coupling to top quarks increases both the gluon fusion production cross section and the loop-induced $H\to \gamma\gamma$ partial width. For $M_H > 2 m_t$ the decay mode $H\to tt$ becomes kinematically accessible, featuring a similar enhancement with $\cot\beta$, and thus the $pp\to H \to \gamma\gamma$ rate does not become large enough to yield an exclusion with current data.

Combining these constraints, in Type I, charged Higgs masses below $\sim 350\gev$ are excluded for $\tb\lesssim2$, while at larger $\tb$ the limit is weaker and arises either from $t\to H^\pm b$, $H^\pm\to \tau\nu$ searches (for $\tb \lesssim 8$) or from LEP searches. Type~II, which resembles the tree-level Higgs sector of the MSSM, is restricted more strongly, i.e.~charged Higgs masses below $\sim 350\gev$ are excluded irrespective of $\tan\beta$.\footnote{Note that in particular for low $M_A$ values around the EW scale, $M_A \sim M_Z$, our assumption $M_{H^+} = M_H = M_A$ is not a good approximation for the MSSM Higgs sector. In fact, the charged Higgs boson mass can still be $M_{H^+}\sim m_t$ in the MSSM --- also consistent with $b\to s\gamma$ measurements --- in the rather exotic scenario where the heavy Higgs boson $H$ is interpreted as the discovered Higgs state~\cite{Bechtle:2016kui}.} At large $\tb \gtrsim 10$, the lower $M_{H^+}$ limit is even more severe due to the $pp\to H/A \to \tau\tau$ constraints, reaching values of $\sim1\tev$ at $\tb \sim 26$. Type~III is slightly less constrained than Type~I at small $\tb \gtrsim 1$, but features an additional excluded region for $\tb\gtrsim20$ obtained from charged Higgs searches and LHC $pp\to H/A\to bb$ searches, constraining the charged Higgs boson to be quite heavy (beyond $800\gev$ in the most stringent case). In Type~IV, the excluded region appears at $\tb\lesssim10$ and $M_{H^+}\lesssim350\gev$. Larger $\tan\beta$ values are not constrained by LHC searches because --- in contrast to Type~II and III --- there is no enhancement of the $H/A$ couplings to bottom quarks.

Flavour physics constraints are complementary to the collider constraints for all four types, as shown by the black dotted contours in Fig.~\ref{2hdm_mAmHmHp_HB}. For Types I and IV, they exclude only the small $\tb$ region, while Higgs searches can reach larger $\tb$ values but smaller charged Higgs masses. For Types II and IV, flavour physics excludes charged Higgs masses larger than $\sim600\gev$ independently of $\tb$. More details and explanations of the flavour physics constraints will be given in Sec.~\ref{sub:flavour}.

Higgs searches allow us to exclude larger charged Higgs masses only for large $\tb$ values due to the $H/A\to\tau\tau$ and $H/A\to b\bar{b}$ constraints (and our assumption $M_{H^+} = M_H = M_A$ in this specific scenario). 

\begin{figure}[t!]
\centering
\includegraphics[width=0.48\columnwidth]{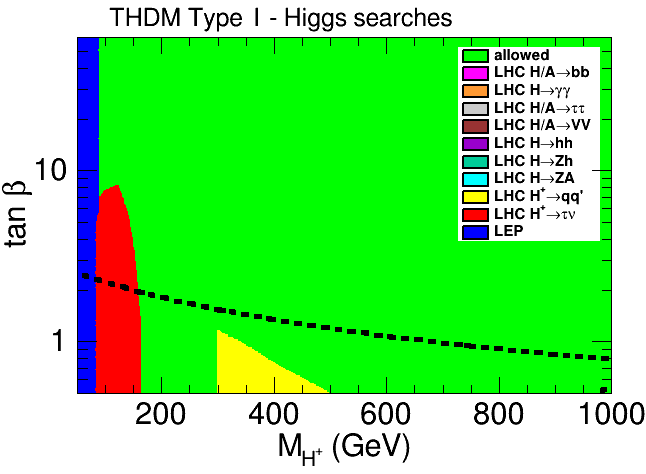}\hfill
\includegraphics[width=0.48\columnwidth]{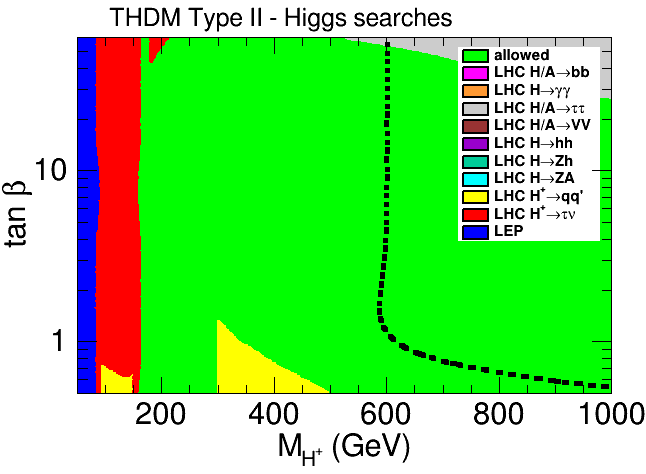}\\
\includegraphics[width=0.48\columnwidth]{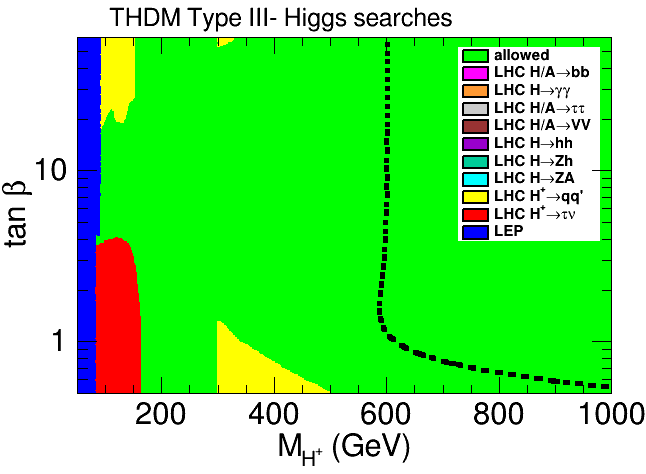}\hfill
\includegraphics[width=0.48\columnwidth]{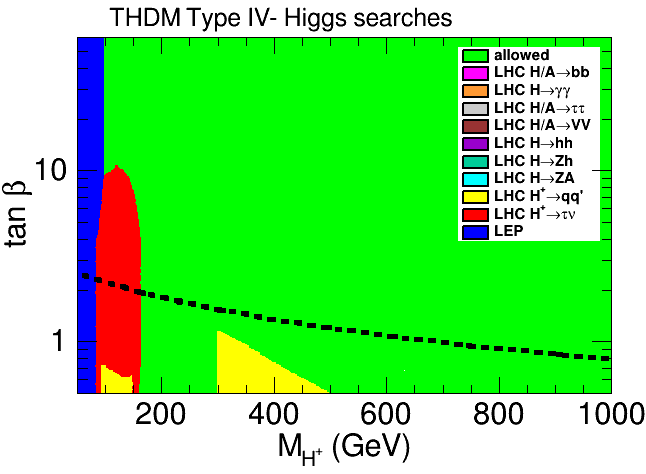}
\caption{Regions of the $(M_{H^+}, \tan\beta)$ parameter space of scenario \textbf{(b)} (\emph{heavy neutral Higgs bosons}) excluded at $95\%$ C.L.\ by charged and neutral Higgs searches (see Sec.~\ref{sec:colliderconstraints}) for the four different 2HDM Yukawa types. The colour coding corresponds to exclusion by different constraints, as given by the legend. The green region is consistent with all collider constraints. The dotted line shows the combined limit from all $b\to s$ observables (see Sec.~\ref{sub:flavour} for details).}
\label{2hdm_mAmH1000_HB}
\end{figure}

In addition, we checked that the constraints on the oblique parameters, $S$, $T$ and $U$, as well as the theoretical bounds from perturbative unitarity and vacuum stability, are satisfied in this scenario.

We now turn to scenario \textbf{(b)} (\emph{heavy neutral Higgs bosons}), where the masses of the heavy neutral Higgs bosons are set to $1\tev$. We show the results for scenario \textbf{(b)} in Fig~\ref{2hdm_mAmH1000_HB}. For all four types, the direct constraints from charged Higgs searches are unaltered with respect to scenario \textbf{(a)}. However, due to the heavy neutral Higgs boson masses, most of the heavy neutral Higgs searches do not provide significant constraints, with the only exception being the $H/A\to\tau\tau$ searches which still give rise to an exclusion for $\tb\gtrsim 30$ and  $M_{H^+}\gtrsim550\gev$ in the Type~II model. At such large $\tb$ values, the $H/A$ couplings to bottom quarks and $\tau$-leptons are significantly enhanced, and the exclusion appears because the branching fraction for $H/A\to\tau\tau$ slightly increases with the charged Higgs mass due to the suppression of the competing $H/A \to W^\pm H^\mp$ decay.
Perturbative unitarity and vacuum stability are fulfilled in this scenario. However, the limits on the oblique parameters impose the strong bound $M_{H^+}\gtrsim900\gev$ independent of $\tb$. Nevertheless, even if most of the presented $(M_{H^+}, \tan\beta)$ parameter plane is disfavoured by the oblique parameters, this scenario is still of interest to illustrate the model-dependence of the neutral Higgs search limits.
The flavour physics constraints are the same as in scenario \textbf{(a)}. Again, these indirect constraints are probing charged Higgs masses far beyond the reach of direct collider searches, and become even more important in cases where the indirect constraints from neutral Higgs searches are irrelevant, because of e.g.~too large Higgs boson masses (as is the case here). In particular in Type~III the flavour physics limits strongly supersede all available limits from Higgs searches.

\begin{figure}[t!]
\centering
\includegraphics[width=0.48\columnwidth]{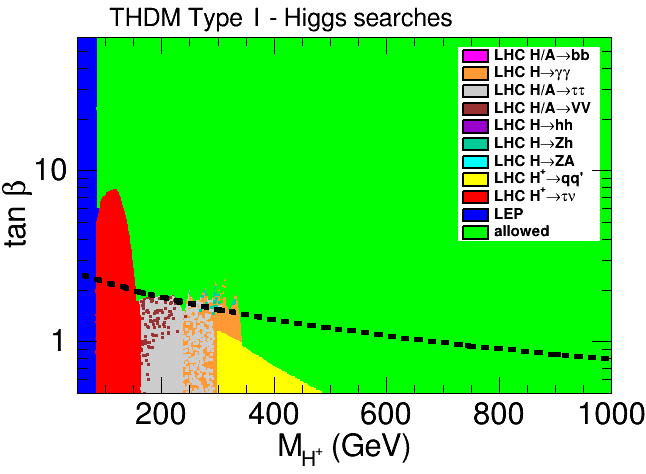}\hfill
\includegraphics[width=0.48\columnwidth]{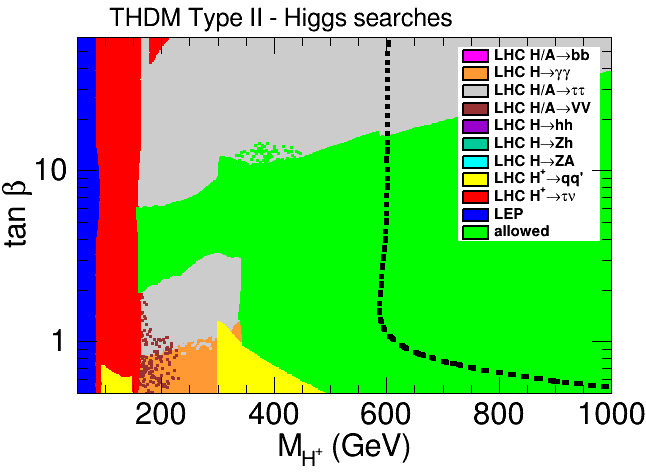}\\
\includegraphics[width=0.48\columnwidth]{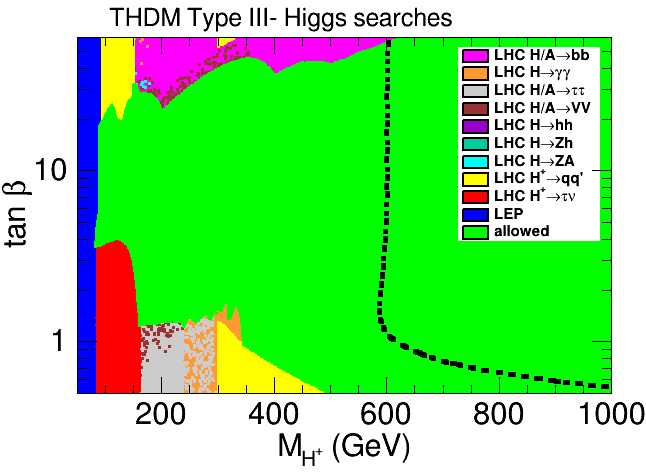}\hfill
\includegraphics[width=0.48\columnwidth]{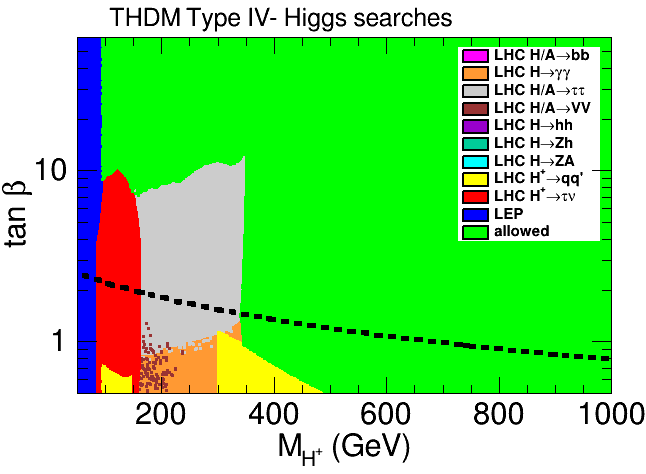}
\caption{Regions of the $(M_{H^+}, \tan\beta)$ parameter space of scenario \textbf{(c)} (\emph{decoupling regime}) excluded at $95\%$ C.L.\ by charged and neutral Higgs searches (see Sec.~\ref{sec:colliderconstraints}) for the four different 2HDM Yukawa types. The colour coding corresponds to exclusion by different constraints, as given by the legend. The points consistent with all collider constraints are shown in the foreground in green. The dotted line shows the combined limit from all $b\to s$ observables (see Sec.~\ref{sub:flavour} for details).}
\label{2hdm_decoup_HB}
\end{figure}

The results for scenario \textbf{(c)} (\emph{decoupling regime}) are shown in Fig.~\ref{2hdm_decoup_HB} in the $(M_{H^+}, \tan\beta)$ parameter plane. In contrast to the previous scenarios, this scenario has three parameters, $M_{H^+}$, $\tan\beta$ and $M_H$. Therefore Fig.~\ref{2hdm_decoup_HB} shows a projection of the parameter space onto the two-dimensional plane $(M_{H^+}, \tan\beta)$. Because of this projection, the order in which the constraints are applied is important. To assess the absolute constraining power of the Higgs searches in this scenario, we plot the points satisfying these constraints at the end (there can therefore be excluded points hidden behind the allowed points). Hence the excluded regions that remain visible are excluded independently of the choice of $M_H$.
We checked that the limits on the oblique parameters and theoretical bounds from perturbative unitarity and vacuum stability do not select specific regions of this parameter plane. We observe again that the constraints from charged Higgs boson searches as well as the flavour physics constraints are unchanged with respect to scenarios \textbf{(a)} and \textbf{(b)}. In contrast, the constraints from neutral Higgs boson searches are slightly weaker than in scenario \textbf{(a)}. This is because in scenario \textbf{(a)} the signal rates from the heavy Higgs bosons $H$ and $A$ are always combined in \texttt{HiggsBounds} when tested against the experimental limits due to the imposed mass degeneracy $M_A = M_H$, whereas in scenario \textbf{(c)} there are always parameter points where $M_A$ and $M_H$ are very different and hence the signal rates are not combined.
In addition, in all Yukawa types, some points at $M_{H^+} \sim 200\gev$ and small $\tb$ are excluded by $H/A\to VV$ ($V = W, Z$) searches. The sensitivity of these searches depends on our assumed scaling behaviour for $\cba$ that emulates the decoupling limit and leads to a small but non-zero $HVV$ coupling at moderately light $M_H$. In Type~III, these searches also rule out points at large $\tb$ in the $M_{H^+}$ range between $160\gev$ and $400\gev$, and additional constraints arise from the $H\to ZA$ searches at $M_{H^+}\sim (170-180)\gev$ and $\tb \sim 30$.

%%%%%%%%%%%%%%%%%%%%%%%%%%%%%%%%%%%%%%%%%%%

\subsubsection{General scenario (d)}

\begin{figure}[th!]
\centering
\vspace*{-0.9cm}\includegraphics[width=0.48\columnwidth]{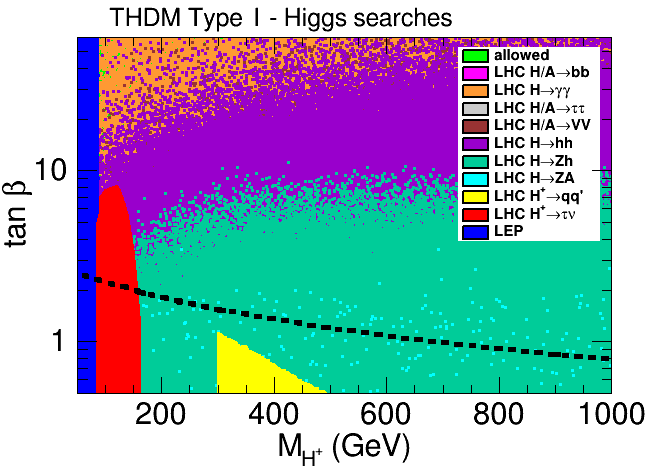} \hfill\includegraphics[width=0.48\columnwidth]{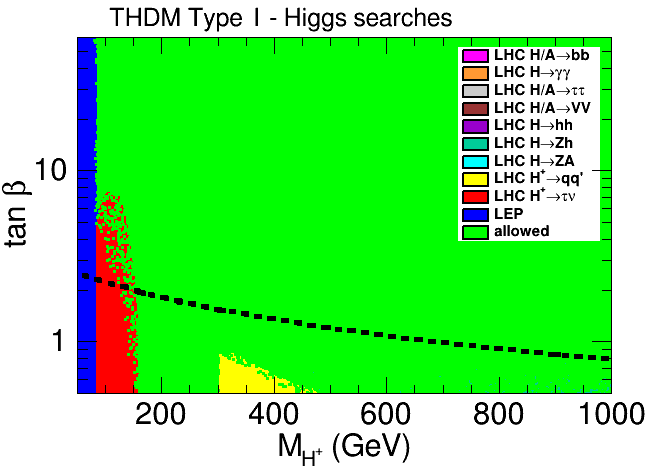}\\[-0.5cm]
\includegraphics[width=0.48\columnwidth]{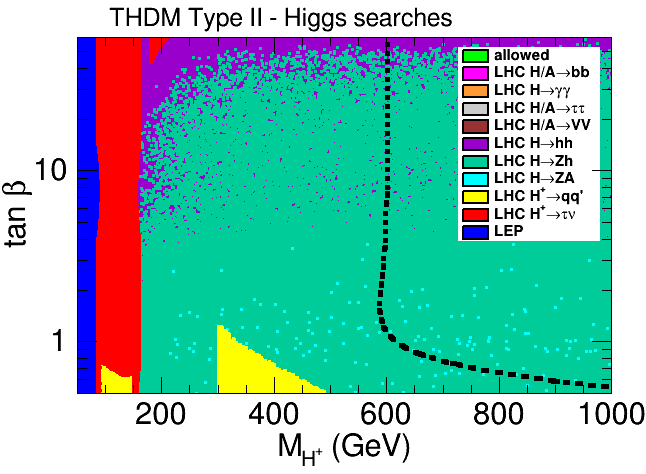} \hfill\includegraphics[width=0.48\columnwidth]{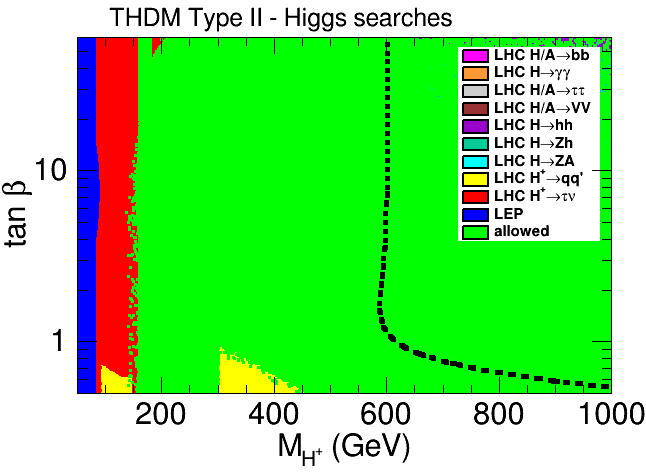}\\[-0.5cm]
\includegraphics[width=0.48\columnwidth]{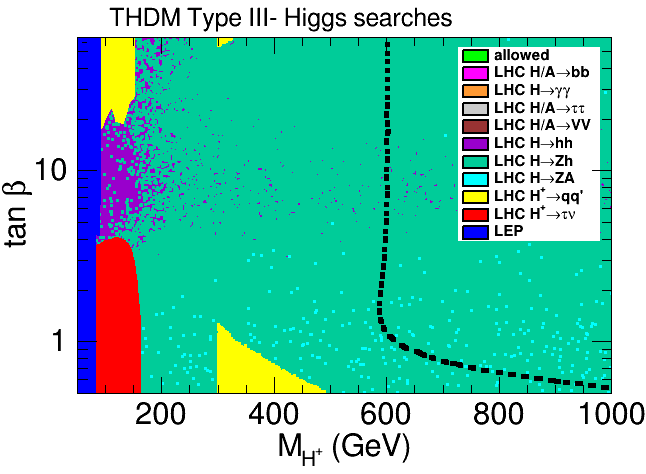} \hfill\includegraphics[width=0.48\columnwidth]{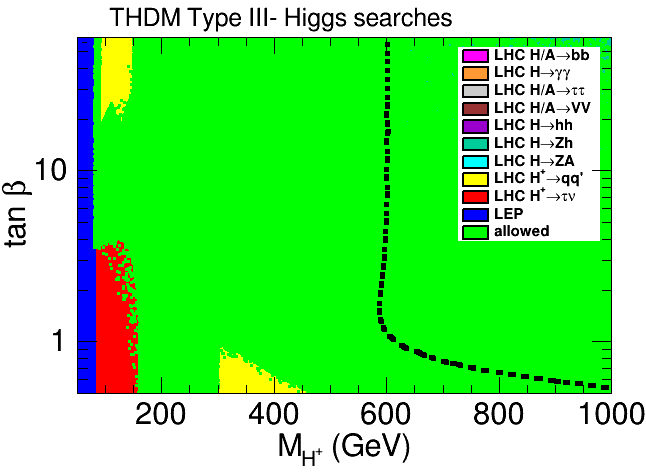}\\[-0.5cm]
\includegraphics[width=0.48\columnwidth]{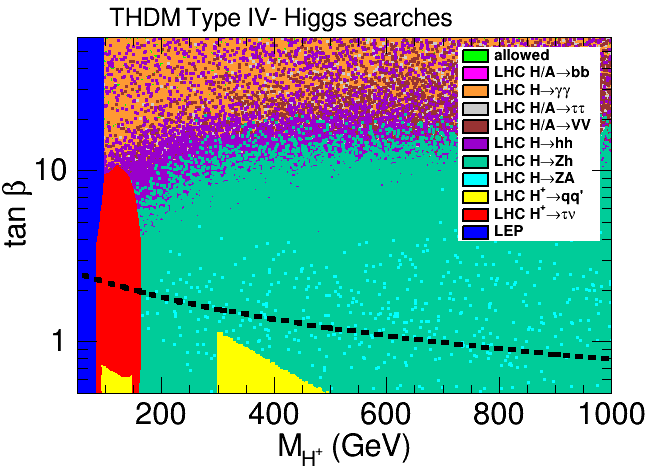} \hfill \includegraphics[width=0.48\columnwidth]{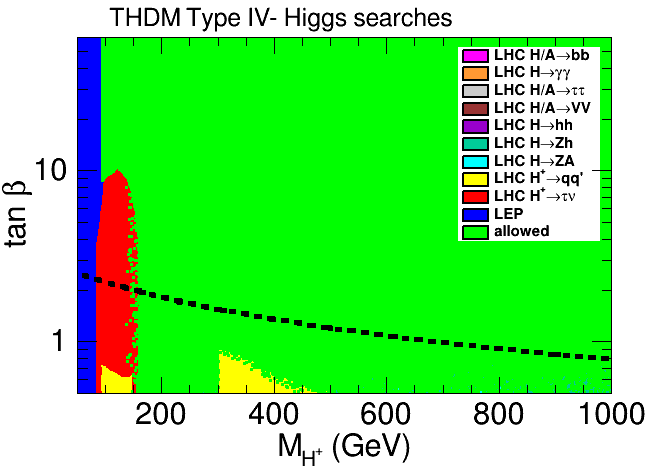}\\[-0.5cm]
\caption{Regions of the $(M_{H^+}, \tan\beta)$ parameter space of scenario \textbf{(d)} (\emph{general scenario}) excluded at $95\%$ C.L.\ by charged and neutral Higgs searches (see Sec.~\ref{sec:colliderconstraints}) for the four different 2HDM Yukawa types. The colour coding corresponds to exclusion by different constraints, as given by the legend. The points consistent with all collider constraints are shown in green in the background in the left panels, and in the foreground in the right column. The dotted line shows the combined limit from all $b\to s$ observables (see Sec.~\ref{sub:flavour} for details).}
\label{2hdm_generic_HB}
\end{figure}

We now investigate scenario \textbf{(d)} (\emph{general scenario}) where we consider the light Higgs to be the discovered Higgs state, $M_h=125.09\gev$, and scan the remaining six parameters, assuming $M_H>m_h$. The results are projected onto the parameter plane $(M_{H^+}, \tan\beta)$, and the order in which the constraints are applied is important. In Fig.~\ref{2hdm_generic_HB} we show in the left (right) panels the results with the allowed points plotted first (last), such that the left (right) panels show the global sensitivity of (robust exclusion from) the various Higgs collider searches.

In the left panels, the excluded regions obtained from charged Higgs searches at LEP and LHC are identical to the regions excluded in the previous scenarios (see Sec.~\ref{sec:results_a-c}). This means that in scenarios \textbf{(a)}-\textbf{(c)} the maximal exclusion of the parameter space is indeed obtained from these searches.
In contrast, we encounter in the general scenario \textbf{(d)} a slight mitigation of the parameter space exclusion obtained from the LHC charged Higgs searches, as can be seen by the green points in the right panels overlapping these areas. For these parameter points the pseudoscalar Higgs mass $M_A$ is small enough such that the charged Higgs decay mode $H^\pm \to A W^{\pm\,*}$ is kinematically accessible.
Additionally, the competing decay mode $H^\pm \to h W^\pm$ could further mitigate these constraints, however, as the responsible $W^\pm H^\pm h$ coupling is proportional to $\cba$, these decays will be suppressed in cases where the light Higgs boson is sufficiently SM-like.

There are several neutral Higgs searches that show sensitivity to this scenario. In addition to the $H/A\to \tau\tau$, $H/A \to bb$ and $H\to \gamma\gamma$ searches, which have been discussed extensively for the previous scenarios \textbf{(a)}-\textbf{(c)}, we find that searches for the Higgs-to-Higgs decays $H\to hh$, $A\to Zh$, $H\to ZA$ and $A\to ZH$ are also sensitive in large regions of the parameter space.\footnote{Parameter points excluded by $H/A\to \tau\tau$, $H/A \to bb$ and $H\to \gamma\gamma$ searches are largely hidden behind the points excluded by Higgs-to-Higgs searches. They are also widely spread over the ($M_{H^+}$, $\tb$) parameter space, since we do not assume any correlation between $M_{H^+}$, $M_H$ and $M_A$ in this scenario.}$^,\,$\footnote{The label ``$H\to ZA$'' for the cyan coloured points in the figures also applies to points excluded by $A\to ZH$ searches.} Of course, these signatures have specific requirements on the Higgs mass spectrum in order to be accessible, and thus can never yield a robust exclusion in the projected ($M_{H^+}$, $\tb$) parameter space unless very specific correlations between the charged and neutral Higgs masses are imposed. Specifically, we find for all 2HDM Yukawa types that the $H\to ZA$ (or $A\to ZH$) exclusions only appear at small $\tb \lesssim 2-3$, whereas the $A\to Zh$ and $H\to hh$ also yield exclusions at larger $\tb$ values (but still with a larger sensitivity towards small $\tb$ values), depending on the Yukawa type.

The oblique parameters and the theoretical bounds strongly reduce the number of valid points, however, in the ($M_{H^+}$, $\tb$) parameter plane, due to the other free parameters, it is always possible to find points compatible with these constraints. In addition, the results from flavour physics in this scenario are unaltered with respect to the previous scenarios \textbf{(a)}-\textbf{(c)}, i.e.~the flavour physics observables are unaffected by the parameters that we allowed to vary additionally here.

%%%%%%%%%%%%%%%%%%%%%%%%%%%%%%%%%%%%%%%%%%%

\subsubsection{Inverted scenario (e)}

\begin{figure}[th!]
\centering
\vspace*{-0.9cm}\includegraphics[width=0.48\columnwidth]{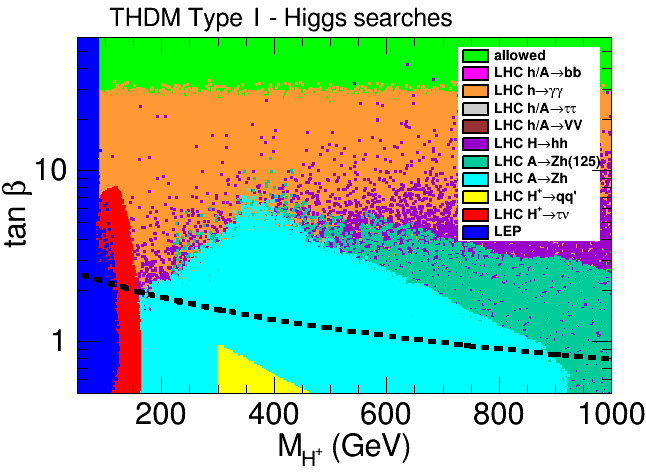}\hfill \includegraphics[width=0.48\columnwidth]{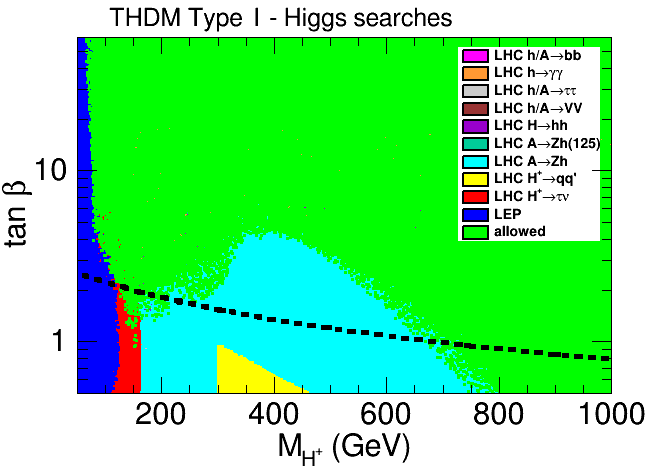}\\[-0.5cm]
\includegraphics[width=0.48\columnwidth]{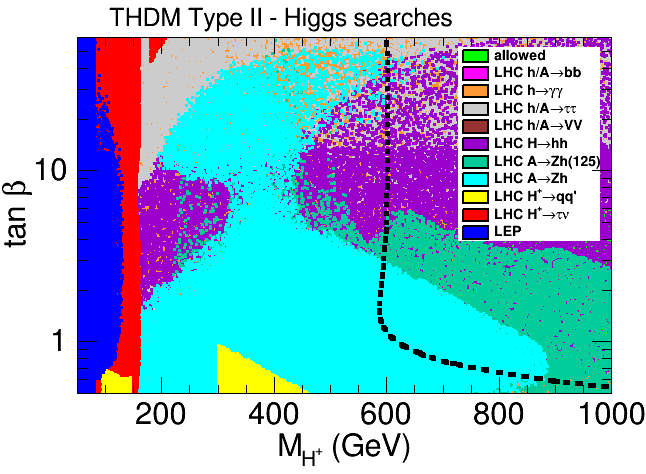}\hfill \includegraphics[width=0.48\columnwidth]{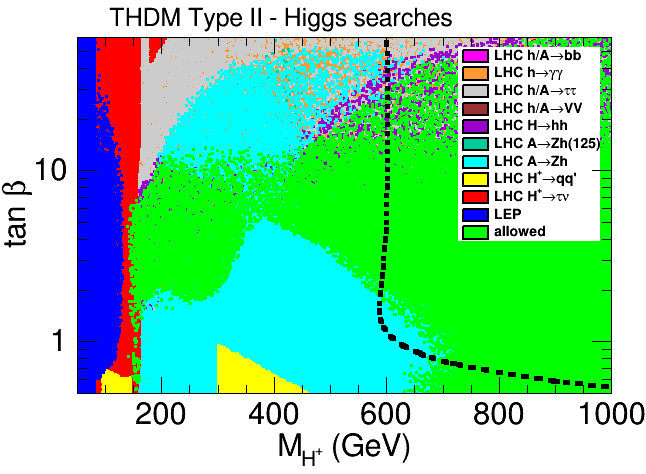}\\[-0.5cm]
\includegraphics[width=0.48\columnwidth]{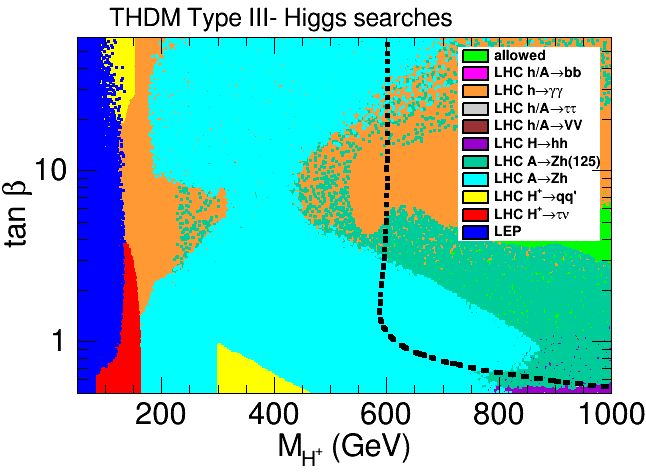}\hfill \includegraphics[width=0.48\columnwidth]{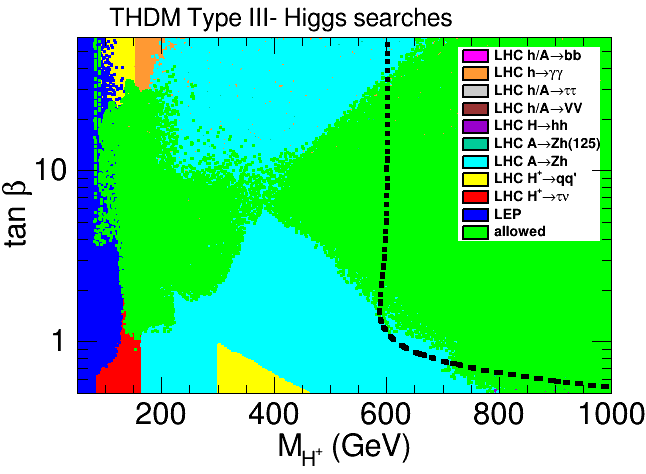}\\[-0.5cm]
\includegraphics[width=0.48\columnwidth]{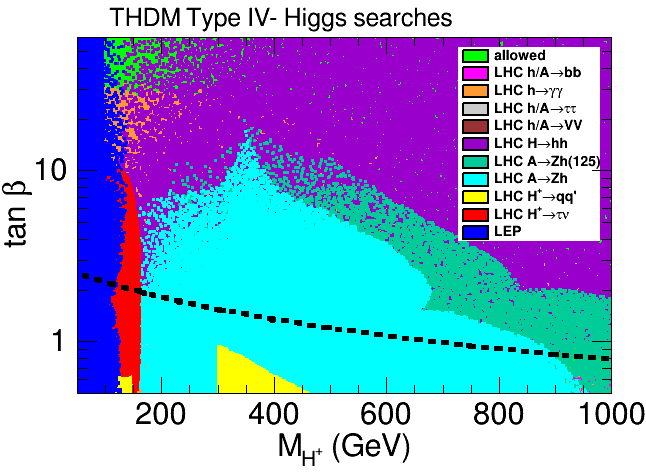}\hfill \includegraphics[width=0.48\columnwidth]{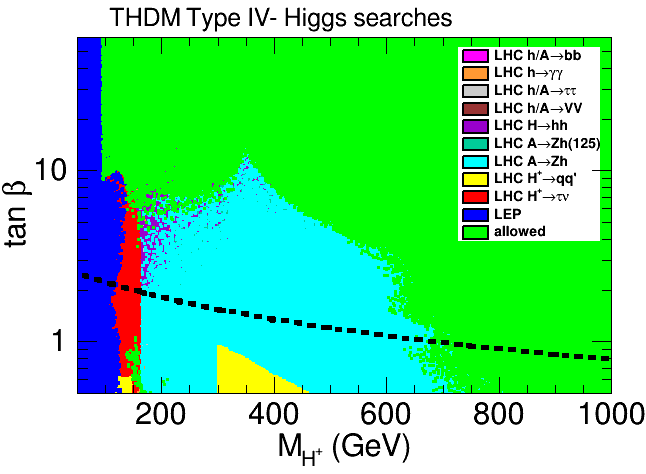}\\[-0.5cm]
\caption{Regions of the $(M_{H^+}, \tan\beta)$ parameter space indirectly excluded by Higgs searches at $95\%$ C.L. for the different 2HDM Yukawa types in the inverted scenario (e) where $M_H=125.09$ GeV and $\cos(\beta-\alpha)=1$. The green region is consistent with all constraints. The allowed points are in the background in the left column, and in the foreground in the right column. The dotted line shows the constraints from all $b\to s$ observables, as explained in section~\ref{sub:flavour}.}
\label{2hdm_invert}
\end{figure}

In scenario \textbf{(e)} (\emph{inverted scenario}) we interpret the discovered Higgs state in terms of the heavy $\CP$-even Higgs boson by setting $M_H = 125.09\gev$. As in the previous scenarios, we show the results in the ($M_{H^+}$, $\tb$) parameter plane in Fig.~\ref{2hdm_invert} for different orders of applying the constraints, i.e.~in the left (right) panels the points allowed by all collider constraints are plotted first (last).

In comparison to the previous scenarios, the excluded region from LEP searches extends to larger $M_{H^+}$ values. Besides the constraints from the combined LEP charged Higgs searches, which lead to identical exclusions in the ($M_H^+$, $\tb$) plane as in the previous scenarios, we have additional constraints from combined LEP searches for the process $e^+e^- \to h A$, with $h$ and $A$ light enough to be kinematically accessible at LEP with center-of-mass energies of up to $\sqrt{s} = 209\gev$, and decaying into pairs of bottom quarks or $\tau$-leptons. Due to this dependence on $M_h$ and $M_A$, however, these excluded regions are not as robust as those obtained from LEP charged Higgs searches, as can be seen by comparing the left and right panels, and furthermore depend strongly on our assumption $M_A = M_{H^+}$ for this scenario. Moreover, note that the assumption $\cba = 1$ made in this scenario also maximises the $ZAh$ coupling that is responsible for the $e^+e^-\to hA$ process.

We find that the parameter space excluded by the LHC charged Higgs searches is similar to what we found for the general scenario \textbf{(d)}. In this scenario, however, the mitigation of the excluded regions only arises from the competing $H^\pm \to  W^\pm h$ decay mode, for which the responsible coupling $\propto \cba$ is maximal in this scenario. We can observe in Fig.~\ref{2hdm_invert} that the exclusions in the light charged Higgs mass region ($M_{H^+}~<~m_t$) are mitigated quite strongly, whereas the exclusion in the heavy charged Higgs mass region ($M_{H^+} > m_t$) is rather unaffected. We will discuss the charged Higgs decay mode $H^\pm \to  W^\pm h$ in more detail below.

Neutral Higgs searches also provide relevant constraints in this scenario. Exclusions from searches for the pseudoscalar Higgs $A$ manifest themselves as distinct areas in the ($M_{H^+}$, $\tb$) parameter space due to our assumption $M_A = M_{H^+}$ in this scenario. Specifically, we find exclusions from $A\to \tau\tau$ and $A\to bb$ searches to be very similar to the ones found in scenario \textbf{(c)} in the ($M_{H^+}$, $\tb$) parameter space. Moreover, due to the presence of a light, non-standard neutral $\CP$-even Higgs boson $h$, we find significant constraints from $A\to Zh$ searches. Here we distinguish two types of searches: first, for a light Higgs boson relatively close to the discovered Higgs state, $M_h\lesssim 125\gev$, searches for $A\to Z h_\text{SM}$ (\emph{dark cyan} points in Fig.~\ref{2hdm_invert}), where the experimental search assumes $h_\text{SM}$ to be the discovered SM-like Higgs boson at $125\gev$;\footnote{The specific requirements on the light Higgs mass depend on the mass resolution of the experimental search and thus on the assumed final state of the decaying SM-like Higgs boson. These are incorporated in an approximate way in \texttt{HiggsBounds}.}
second, searches for $A\to Zh$ (\emph{light cyan} points in Fig.~\ref{2hdm_invert}), where $h$ is a non-standard Higgs boson. For the latter type of searches cross section limits are available for the entire $M_A$ and $M_h$ scan ranges considered here in this scenario. For all Yukawa types, both types of $A\to Zh$ searches yield exclusions at small to moderate $\tb$ values, and are strongest near the kinematic threshold of the $A\to tt$ decay, i.e.\ at $M_{H^+} = M_A \sim 350\gev$. In addition, in the Type II and Type III models, they also exclude parts of the parameter space at large $\tb$ due to the enhanced $A$ production in association with bottom quarks. Remarkably, a large portion of the exclusion from $A\to Zh$ searches (with non-standard Higgs boson $h$, \emph{light cyan}) found in the left panels is robust under variation of the remaining parameters (in particular, $M_h$), as can be seen when comparing with the right panels. Note, that the decay $A\to ZH$ (where $H$ is the SM-like Higgs boson) does not provide constraints in this scenario as the $ZAH$ coupling is $\propto \sba$ and thus vanishes in the exact alignment limit that we assume here.

If $M_h$ is small enough, $M_h \lesssim 62.5\gev$, the decay $H\to hh$ is in general possible and leads to excluded points for all Yukawa types, albeit with a large variation of the distribution of the points among the different types in the ($M_{H^+}$, $\tb$) parameter space, as can be seen in the left panels in Fig.~\ref{2hdm_invert}. However, the exclusions from these searches are in principle independent of $M_{H^+}$ (and $M_A$). The shape of the point distribution in the left panels of Fig.~\ref{2hdm_invert} for these constraints are either due to an overlay of other constraints or due to a competition with other search channels in being the most sensitive channel selected by \texttt{HiggsBounds}. Note, however, that the $H\to hh$ decay rates cannot be too large as they would otherwise spoil the compatibility with the measured Higgs rates.

\begin{figure}[t!]
\centering
\vspace*{-0.5cm}\includegraphics[width=0.48\columnwidth]{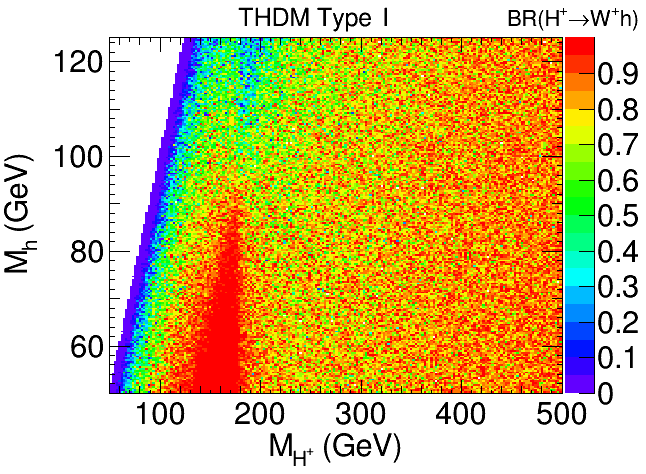}\hfill \includegraphics[width=0.48\columnwidth]{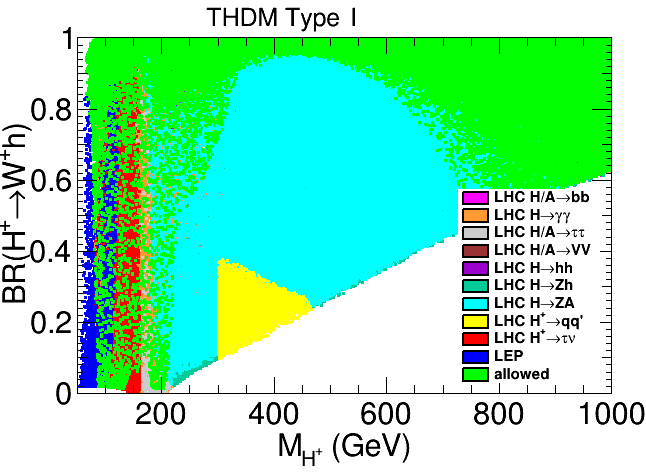}\\[-0.5cm]
\includegraphics[width=0.48\columnwidth]{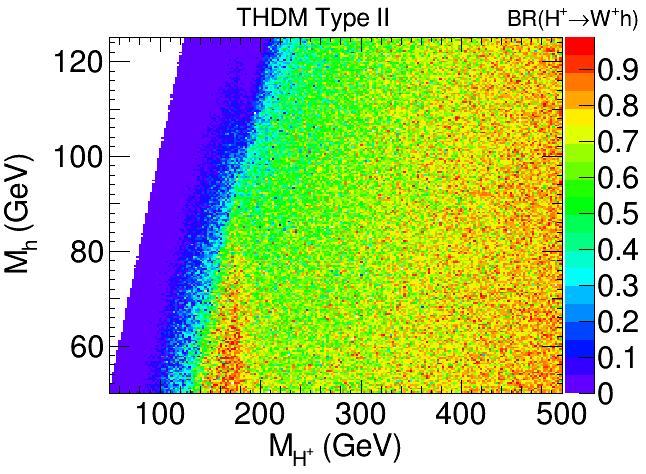}\hfill \includegraphics[width=0.48\columnwidth]{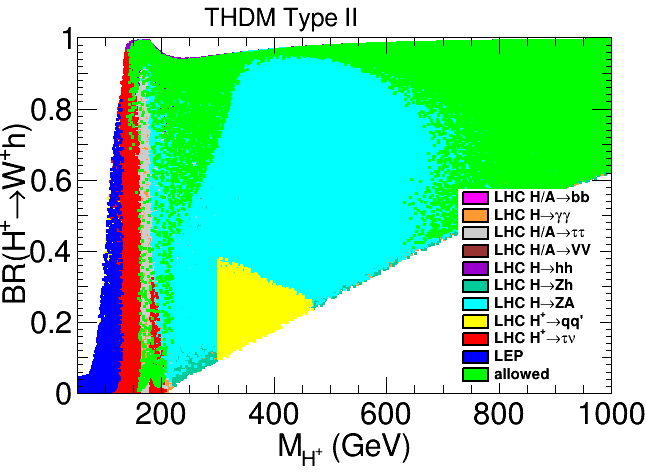}\\[-0.5cm]
\includegraphics[width=0.48\columnwidth]{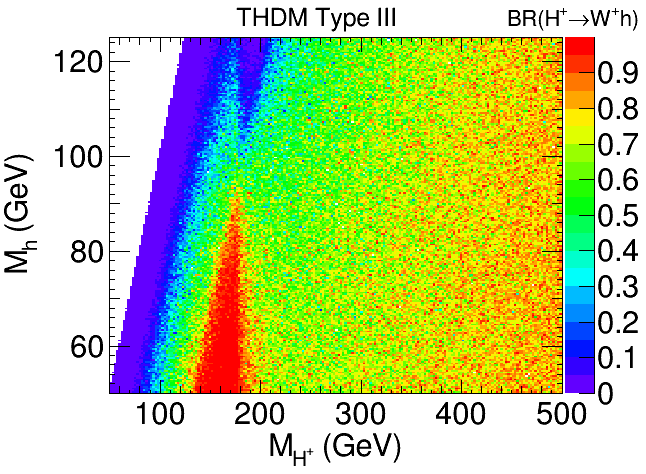}\hfill \includegraphics[width=0.48\columnwidth]{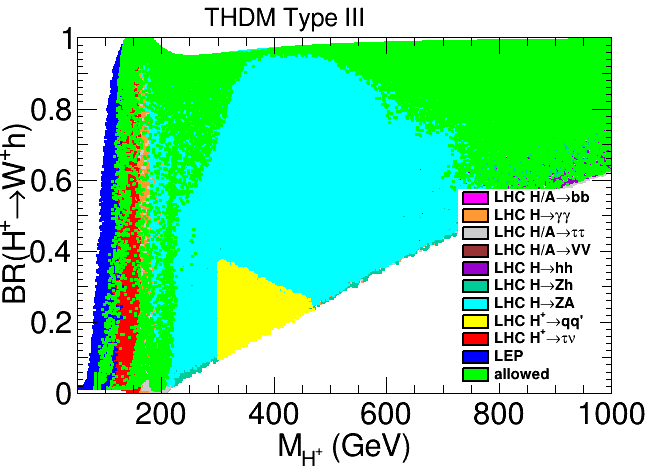}\\[-0.5cm]
\includegraphics[width=0.48\columnwidth]{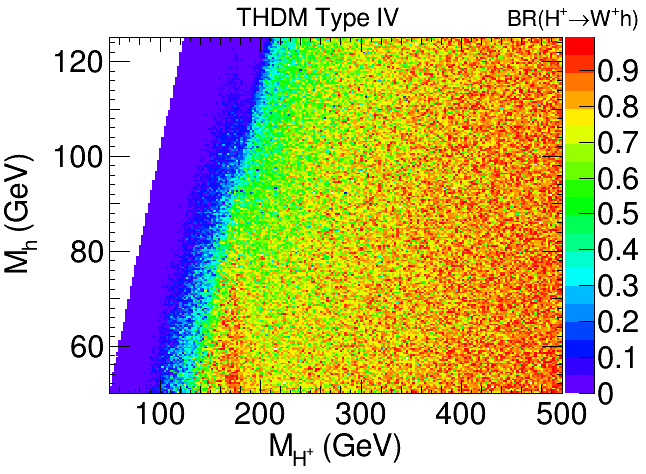}\hfill \includegraphics[width=0.48\columnwidth]{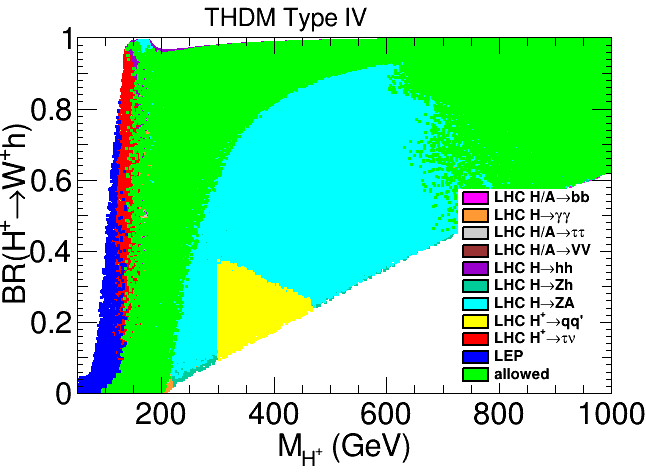}\\[-0.4cm]
\caption{Branching fraction of the charged Higgs boson decay $H^\pm \to W^\pm h$ in the ($M_{H^+}$, $M_h$) parameter plane (\emph{left panels}) and  as a function of $M_{H^+}$ (\emph{right panels}), for the different 2HDM Yukawa types in the inverted scenario \textbf{(e)} where $M_H=125.09$ GeV and $\cos(\beta-\alpha)=1$. The left panels contain all scan points, with larger values plotted on top of small values. The color coding in the right panels is identical to the right panels of Fig.~\ref{2hdm_invert}, i.e.~the allowed points are plotted last.}
\label{2hdm_invert_HpWh}
\end{figure}

Additional constraints also arise from LHC searches for $h\to \gamma\gamma$, for which limits are available for $M_h \ge 65\gev$. These searches exhibit sensitivity in almost the entire ($M_{H^+}$, $\tb$) parameter plane as there is only a mild connection with $M_{H^+}$.\footnote{This mild connection comes from a possible enhancement of the $h\to \gamma\gamma$ decay due to the loop contribution of a light charged Higgs boson. However, this effect seems to have negligible impact on our results.}

We note that perturbativity, and to a lesser extent unitarity, favour points with $M_{H^+}\lesssim600\gev$ and $\tb\lesssim10$.

We now come back to the discussion of the charged Higgs boson decay $H^\pm \to W^\pm h$. Fig.~\ref{2hdm_invert_HpWh} shows the branching fraction $\mathrm{BR}(H^\pm \to W^\pm h)$ in the ($M_{H^+}$, $M_h$) parameter plane (left panels), and as a function of $M_{H^+}$ (right panels). For all Yukawa types, we find that $\mathrm{BR}(H^\pm \to W^\pm h)$ can quite generically become very large ($\lesssim 100\%$) in nearly the whole ($M_{H^+}$, $M_h$) parameter plane except for the case where $\Delta M \equiv M_{H^+} - M_h $ is smaller than around $(20, 70, 40, 70)\gev$ in the Type (I, II, III, IV) models, respectively. In addition, in Type III, the region $M_{H^+} \sim (180 - 220)\gev$ and $M_h \sim (100 - 125)\gev$ also exhibits low $H^\pm \to W^\pm h$ decay rates. An accumulation of large values, $\mathrm{BR}(H^\pm \to W^\pm h) \lesssim 100\%$, can be identified in the interesting low to intermediate charged Higgs mass regime, $M_{H^+} \sim (120 - 200)\gev$, with light Higgs boson masses $M_h \lesssim (80 - 100)\gev$ (exact values depending on the Yukawa type, see Fig.~\ref{2hdm_invert_HpWh}). Large decay rates are also found quite generically for very heavy charged Higgs bosons.

The right panels of Fig.~\ref{2hdm_invert_HpWh} use the same color coding for the excluded points as in the right panels of Fig.~\ref{2hdm_invert}, i.e.~the parameter points that pass all the collider constraints are plotted at the end. We can make two important observations here: first, in the light charged Higgs regime ($M_{H^+} < m_t$), the direct charged Higgs searches at LEP and LHC are still capable of excluding parameter points with quite large $\mathrm{BR}(H^\pm \to W^\pm h)$, although they focussed mostly on the $\tau\nu$ and $cs$ final states (see Sec.~\ref{sub:chargedcollider}); second, in the heavy charged Higgs regime, $M_{H^+} \sim (220 - 700)\gev$, the $A\to Zh$ (with a non-standard light Higgs boson $h$) excludes the low values of $\mathrm{BR}(H^\pm \to W^\pm h)$. This latter observation illustrates a remarkable complementarity of the neutral Higgs search for $A\to Zh$ and a possible future charged Higgs search for $pp \to H^\pm tb$ with $H^\pm \to W^\pm h$ in this mass regime. In particular, in the range $M_{H^+} \sim (300 - 650)\gev$, the decay $H^\pm \to W^\pm h$ is always dominating ($\gtrsim 50\%$) for the allowed points in this scenario. In conclusion, charged Higgs searches for the decay $H^\pm \to W^\pm h$ form a promising future experimental avenue that complements the currently existing charged and neutral Higgs searches at the LHC. For further discussion of the collider search prospects for this signature see Refs.~\cite{Coleppa:2014cca,Kling:2015uba,Kling:2016opi}.

%%%%%%%%%%%%%%%%%%%%%%%%%%%%%%%%%%%%%%%%%%%

\subsection{Higgs rate measurements}
\label{sec:Higgsrates}

We now turn to the constraints from Higgs rate measurements. In the alignment limit, $\sba \to 1$, the light Higgs state obtains tree-level couplings to fermions and gauge bosons as in the SM, and is therefore expected to be consistent with the LHC Higgs rate measurements. It is however interesting to see whether a charged Higgs boson can influence this picture significantly, and in turn, whether one can derive constraints on the charged Higgs sector from the Higgs rate measurements.

\begin{figure}[t!]
\centering
\includegraphics[width=0.48\columnwidth]{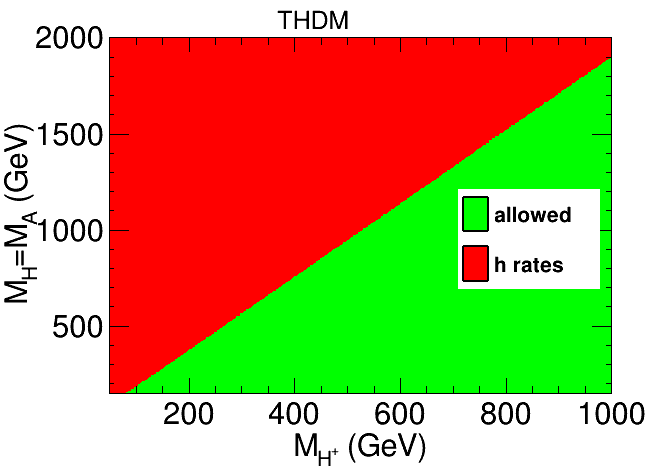}\hfill \includegraphics[width=0.48\columnwidth]{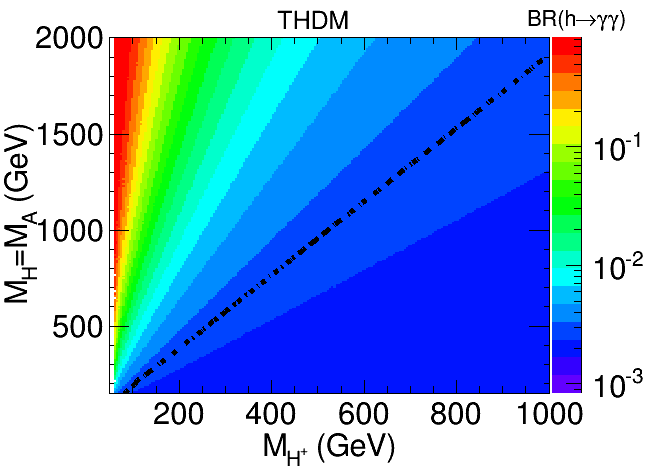}\\
\caption{\emph{Left:} Regions of the $(M_{H^+}, M_H=M_A)$ parameter space indirectly excluded by the Higgs rate measurements at $95\%$ C.L. for the different 2HDM Yukawa types. \emph{Right:} Branching ratio of $h\to \gamma\gamma$ in the $(M_{H^+}, M_H=M_A)$ parameter plane. The dashed line delimits the exclusion by the Higgs rate measurements.}
\label{2hdm_rates}
\end{figure}%
Here, we consider a scenario with $M_h=125.09\gev$, $\sba=1$, $\tb=10$ and $m_{12}^2=M_A^2 \tb/(1+\tan^2\beta)$. We allow $M_H=M_A$ and $M_{H^+}$ to vary independently. All four Yukawa types yield identical results, which are shown in Fig.~\ref{2hdm_rates}. In the left panel, the exclusion by the Higgs rate measurements is shown in the $(M_{H^+}, M_H=M_A)$ parameter plane. In contrast to the intuitive idea that the alignment limit leads to an automatic agreement with the Higgs rate measurements, we find that a large part of the parameter plane is excluded. In fact, this is due to the enhancement of the $h\to\gamma\gamma$ decay rate via the charged Higgs boson loop. These contributions increase with the $hH^\pm H^\mp$ coupling, which in turn increases with $M_H$ and $M_A$. This dependence is illustrated in the right panel, where we show the branching ratio $\mathrm{BR}(h\to\gamma\gamma)$ in the ($M_{H^+}, M_{H}=M_{A}$) parameter plane. At large $M_{H}=M_{A}$ values and small $M_{H^+}$ values we find very large enhancements of the branching fraction of $h\to\gamma\gamma$ in comparison to its SM value, and thus this parameter region turns out to be inconsistent with the Higgs rate measurements. The obtained exclusion line is diagonal in this parameter plane.

However, using more general scans, we checked that the constraints on the charged Higgs mass can be easily circumvented by changing other parameters that directly affect the light Higgs phenomenology, in particular, the mixing angle $\alpha$. In that case, even though many parameter points will still be excluded by the Higgs rates, allowed points can be found in the entire parameter plane displayed in Fig.~\ref{2hdm_rates}.

%%%%%%%%%%%%%%%%%%%%%%%%%%%%%%%%%%%%%%%%%%%

\subsection{Flavour physics}
\label{sub:flavour}

Similarly to the direct constraints from colliders, some of the constraints on the charged Higgs parameters from flavour physics depend strongly on the 2HDM Yukawa type, while others are type-independent. The combined flavour constraints are shown in Figs.~\ref{2hdm_flav1} and \ref{2hdm_flav2} for the four Yukawa types, where the different colours show the parameter regions excluded at 95\% C.L.\ by different flavour observables (as indicated in the legend, see Sec.~\ref{sec:flav-constraints} and Ref.~\cite{Mahmoudi:2009zx} for more details). The displayed results have been obtained within the general scenario \textbf{(d)}, however, all the scenarios defined in Sec.~\ref{sec:2hdmscenarios} lead to almost identical results for each of the $Z_2$-symmetric types, and therefore the obtained exclusions are essentially scenario-independent. This is due to the fact that $M_{H^+}$ and $\tan\beta$ are the two parameters that strongly dominate the flavour observables. Therefore, in contrast to the supersymmetric case, in the 2HDM these constraints have a negligible dependence on the remaining parameters of the model. A remarkable feature that we can immediately observe from the figures is that the $M_{H^\pm}$ range accessible to indirect constraints from flavour physics extends far beyond what is probed by direct collider searches. In the type II and III models, a $\tb$-independent lower limit of $M_{H^\pm}\gtrsim 600\gev$ is obtained from the $\mathrm{BR}(B\to X_s\gamma)$ constraint.

\begin{figure}[t!]
\centering
\includegraphics[width=0.48\columnwidth]{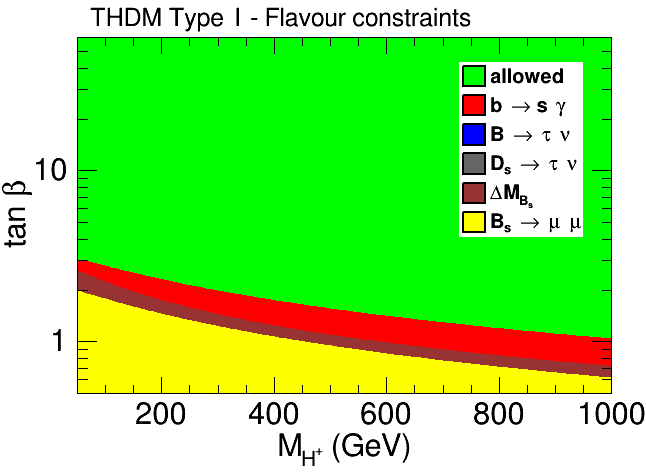}\hfill
\includegraphics[width=0.48\columnwidth]{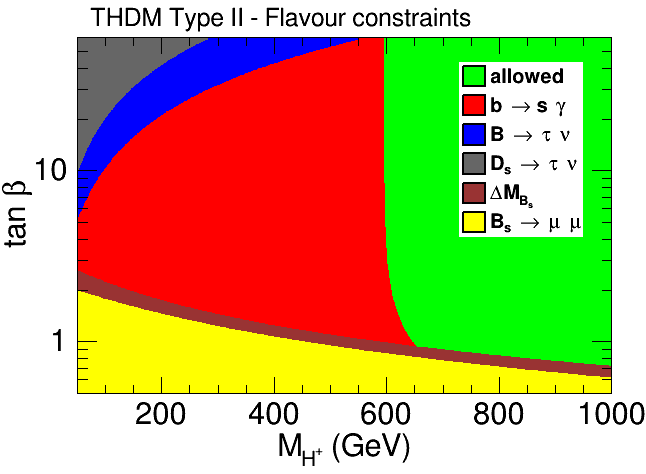}\\
\includegraphics[width=0.48\columnwidth]{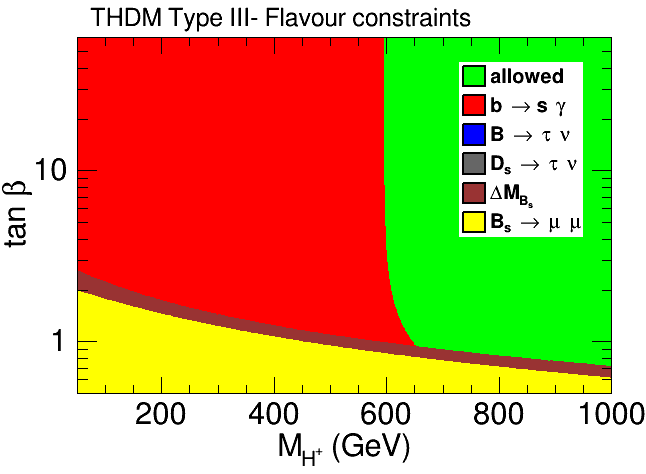}\hfill
\includegraphics[width=0.48\columnwidth]{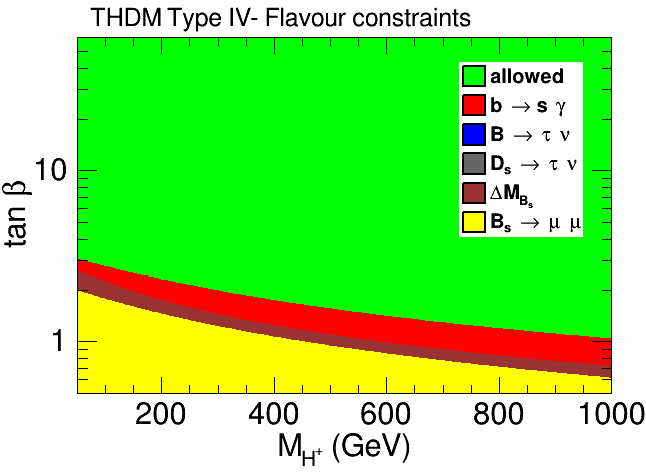}
\caption{Regions of the $(M_{H^+}, \tan\beta)$ parameter space excluded by flavour constraints at $95\%$ C.L. for the different 2HDM Yukawa types. The colour coding corresponds to exclusion by different constraints, as described by the legend. The green region is consistent with all constraints. The obtained constraints are independent of the remaining 2HDM parameters. }
\label{2hdm_flav1}
\end{figure}

At low $\tb$ the constraints are similar among all Yukawa types as the Higgs couplings to up-type quarks are universal. For all types, we find that values of $\tb < 1$ are ruled out for all values of $M_{H^\pm} < 650$ GeV, which includes the interesting non-decoupling region of the type I model. There are several observables which probe mainly the type II model at high $\tb$, with the most sensitive one being the $\mathrm{BR}(B\to \tau\nu)$.
There is a correlation between this mode (and several other similar leptonic decays) and direct collider searches for $H^\pm\to \tau\nu$, since the same couplings enter the dominant 2HDM contribution.

The very strong constraint from the branching ratio $\mathrm{BR}(B\to X_s\gamma)$ is due to the contribution from the charged Higgs boson, which enters at the same level as the $W$ boson contribution in the SM~\cite{Misiak:2017bgg}. The Wilson coefficient $C_7$ depends on the Yukawa couplings $\lambda_{tt}$ and $\lambda_{bb}$. Since the Yukawa types II and III share the same coupling pattern for the quarks, $|\lambda_{tt}|=1/|\lambda_{bb}|$, one obtains a $\tb$-independent lower limit on the charged Higgs mass for these two types from $\mathrm{BR}(B\to X_s\gamma)$.

The decay $B_s\to \mu^+\mu^-$, which is very constraining in supersymmetric models~\cite{Akeroyd:2011kd,Arbey:2012ax}, is sensitive to large values of $\lambda_{\mu\mu}$, $\lambda_{bb}$ and $\lambda_{tt}$, and small charged Higgs masses. In the 2HDM, it constrains therefore mainly the low $\tb$ region.

For the decay $B_u\to \tau\nu$ (and similarly $D_s\to \tau\nu$) the charged Higgs boson appears already at tree-level, and does not suffer from the helicity suppression as does the $W$ boson in the SM. The interference, which is proportional to the product $\lambda_{bb} \lambda_{\tau\tau}$, is therefore rather large and leads to a reduced branching fraction. In type II, $\lambda_{bb} \lambda_{\tau\tau} \sim \tan^2\beta$, and hence $B_u \to \tau\nu$ leads to strong constraints at large $\tb$.

\begin{figure}[t!]
\centering
\includegraphics[width=0.48\columnwidth]{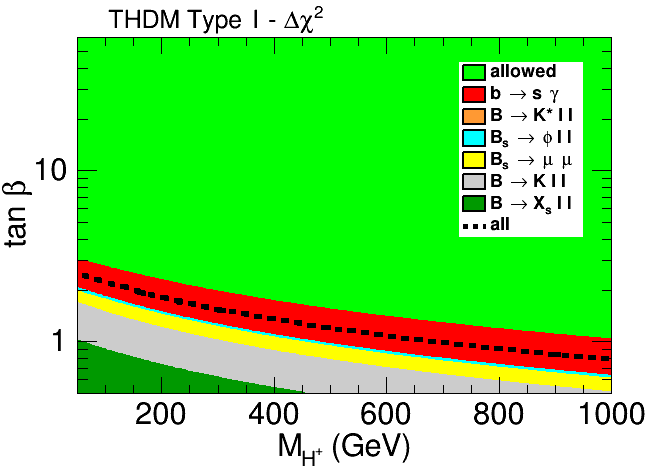}\hfill
\includegraphics[width=0.48\columnwidth]{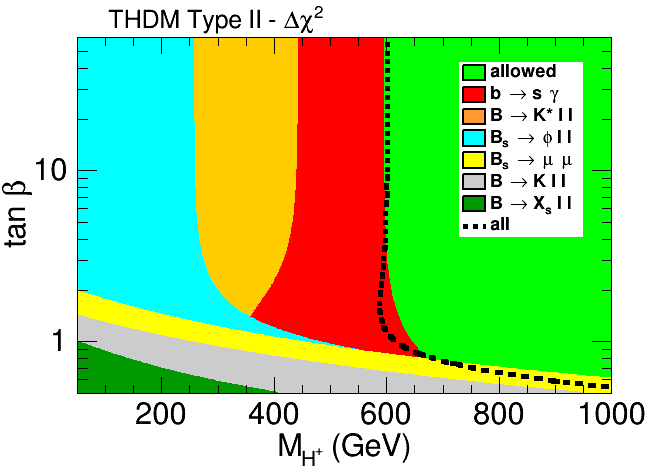}\\
\includegraphics[width=0.48\columnwidth]{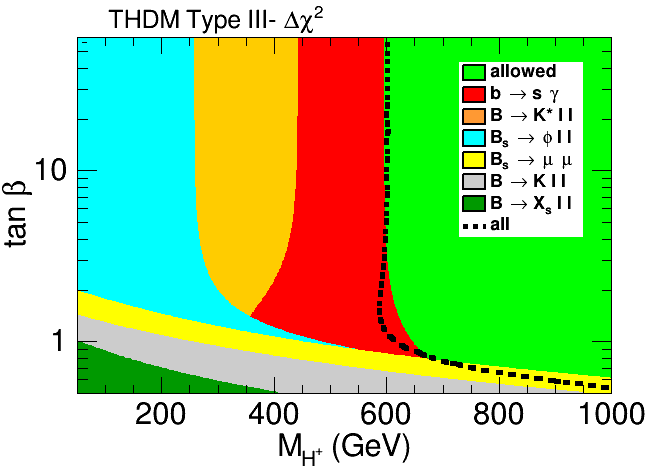}\hfill
\includegraphics[width=0.48\columnwidth]{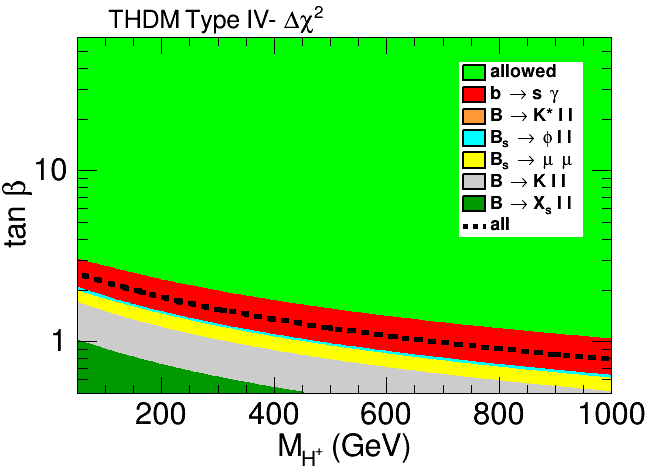}
\caption{Regions of the $(M_{H^+}, \tan\beta)$ parameter space excluded by flavour constraints at $95\%$ C.L. for the different 2HDM Yukawa types. The colour coding corresponds to exclusion by different constraints, as described by the legend. The green region is consistent with all constraints.}
\label{2hdm_flav2}
\end{figure}

In Fig.~\ref{2hdm_flav2}, we show in addition the constraints from $B\to K \ell^+\ell^-$ branching ratios, $B\to K^{*}\ell^+\ell^-$ branching ratios and angular observables, $B_s\to \phi\mu^+\mu^-$ branching ratio and angular observables and $B\to X_s \ell^+\ell^-$ branching ratios and forward-backward asymmetries. The full list of the employed observables can be found in Ref.~\cite{Hurth:2016fbr}. The combined constraints obtained based on a $\chi^2$ study (see Sec.~\ref{sec:flav-constraints}) is also provided. While the very strong constraint from BR($B\to X_s\gamma$) is dominant for the four types, its consistency and complementarity with these new observables is certainly interesting. For comparison, the combined flavour constraints shown by the dashed curves in the Fig.~\ref{2hdm_flav2} are also plotted for each scenario and each type in the figures showing the direct LHC Higgs search constraints in Sec.~\ref{sec:lhc}. This clearly illustrates that these limits are very robust and independent of the specific parameter choices made in our different scenarios.

%%%%%%%%%%%%%%%%%%%%%%%%%%%%%%%%%%%%%%%%%%%

\subsection{2HDM with generic couplings and $B \to D^{(*)} \ell \nu$}

We now turn to the constraints from $B \to D^{(*)} \ell \nu$ transitions. Specifically, we study the implications of the constraints given in Table~\ref{ObsBDlnu} at 95\% C.L for the 2HDM. None of the above $Z_2$-symmetric 2HDM scenarios can provide parameter points consistent with these constraints. We therefore investigate this problem in scenario \textbf{(f)}, where the charged Higgs couplings are considered to be generic. In particular, $B \to D^{(*)} \ell \nu$ transitions are sensitive to $\lambda_{bb}$, $\lambda_{cc}$, $\lambda_{\tau\tau}$, $\lambda_{ee}$ and $\lambda_{\mu\mu}$ and the charged Higgs mass \cite{Tanaka:2012nw,Sakaki:2013bfa}. Since the $\lambda_{ee}$ and $\lambda_{\mu\mu}$ terms are multiplied by the electron and muon masses, respectively, they are strongly suppressed, and therefore are irrelevant. More precisely, three parameter combinations appear in the equations: $\lambda_{bb}\lambda_{\tau\tau}$, $\lambda_{cc}\lambda_{\tau\tau}$ and $M_{H^+}$. Allowing these parameters to vary freely, in addition to the other ``standard'' 2HDM parameters (see Sec.~\ref{sec:thdm} for details), we find parameter points in agreement with the data at the 95\% C.L. We present the results in Fig.~\ref{2hdm_bdlnu}, in the ($\lambda_{cc}\lambda_{\tau\tau}/\sqrt{|\lambda_{cc}\lambda_{\tau\tau}|},\lambda_{bb}\lambda_{\tau\tau}/\sqrt{|\lambda_{bb}\lambda_{\tau\tau}|}$) and ($M_{H^+},\lambda_{bb}\lambda_{\tau\tau}/\sqrt{|\lambda_{bb}\lambda_{\tau\tau}|}$) parameter planes. The normalisations have been chosen so that they can be directly compared to the Type I-IV couplings. The correspondence is given in Table~\ref{tab:gentypecorresp}.

\begin{figure}[t!]
\centering
\includegraphics[width=0.48\columnwidth]{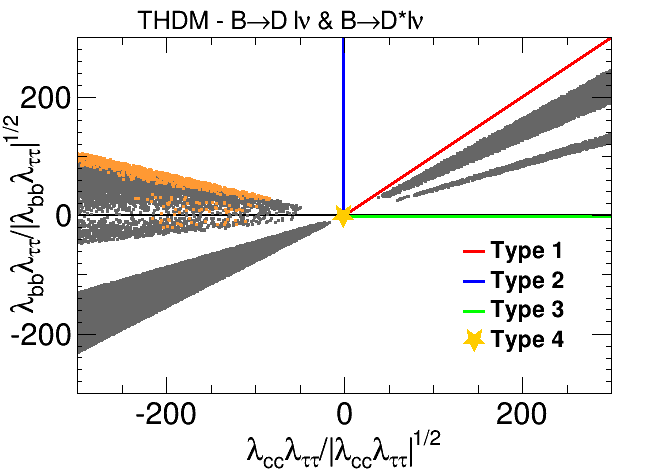}\hfill
\includegraphics[width=0.48\columnwidth]{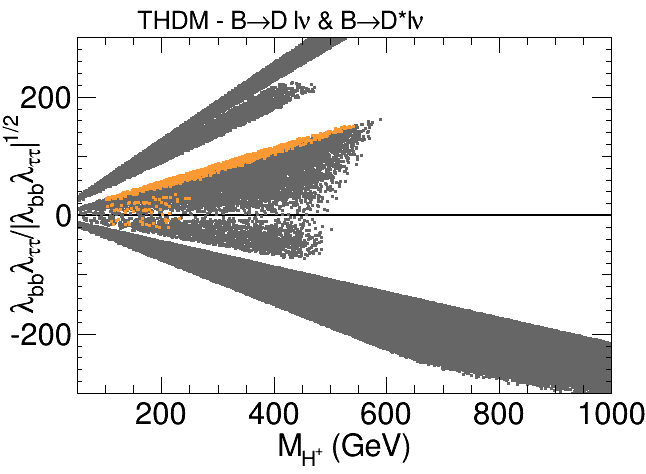}
\caption{Results in the ($\lambda_{cc}\lambda_{\tau\tau}/\sqrt{|\lambda_{cc}\lambda_{\tau\tau}|},\lambda_{bb}\lambda_{\tau\tau}/\sqrt{|\lambda_{bb}\lambda_{\tau\tau}|}$) (\emph{left}) and the ($M_{H^+},\lambda_{bb}\lambda_{\tau\tau}/\sqrt{|\lambda_{bb}\lambda_{\tau\tau}|}$) (\emph{right}) parameter plane for scenario \textbf{(f)} (\emph{generic couplings}). The grey points are in agreement with $B \to D^{(*)} \ell \nu$ constraints at $95\%$ C.L., and the orange points are in addition in agreement with $\mathrm{BR}(B_u\to\tau\nu)$ and $\mathrm{BR}(D_s\to\tau\nu)$. The lines and star correspond to the regions accessible for the various $Z_2$-symmetric Yukawa types.\label{2hdm_bdlnu}}
\end{figure}

\begin{table}[t!]
\centering
\begin{tabular*}{0.7\columnwidth}{@{\extracolsep{\fill}}ccc}
\hline
Type & $\lambda_{cc}\lambda_{\tau\tau}/\sqrt{|\lambda_{cc}\lambda_{\tau\tau}|}$ & $\lambda_{bb}\lambda_{\tau\tau}/\sqrt{|\lambda_{bb}\lambda_{\tau\tau}|}$ \\
\hline
I & $\cot\beta$ & $\cot\beta$\\
II & $-1$ & $\tan\beta$\\
III & $\cot\beta$ & $-1$\\
IV & $-1$ & $-1$\\
\hline
\end{tabular*}
\caption{Values of the parameters entering $B \to D^{(*)} \ell \nu$ observables for the different $Z_2$-symmetric Types.\label{tab:gentypecorresp}}
\end{table}

We can identify in Fig.~\ref{2hdm_bdlnu} four different branches of grey points which respect the $B \to D^{(*)} \ell \nu$ constraints, two of them with both $\lambda_{cc}\lambda_{\tau\tau}$ and $\lambda_{bb}\lambda_{\tau\tau}$ positive, and two with negative $\lambda_{cc}\lambda_{\tau\tau}$. However, $\lambda_{bb}\lambda_{\tau\tau}$ is also constrained by $\mathrm{BR}(B_u\to\tau\nu)$ and $\lambda_{cc}\lambda_{\tau\tau}$ by $\mathrm{BR}(D_s\to\tau\nu)$. Imposing simultaneously the $B \to D^{(*)} \ell \nu$, $B_u\to\tau\nu$ and $D_s\to\tau\nu$ constraints at the 95\% C.L.\ reduces the number of allowed points significantly. These are given by the orange points, which mostly form a small strip with $\lambda_{cc}\lambda_{\tau\tau}$ negative. We further observe that the charged Higgs mass is limited to be below $550\gev$. These points have $|\lambda_{bb}|\sim\mathcal{O}(10)$, whereas $|\lambda_{\tau\tau}|$ and $|\lambda_{cc}|\sim\mathcal{O}(100)$. In addition, the other couplings can be adjusted in order to reach agreement with the other flavour constraints used in the previous section. Finally, we have checked that most of these orange points are also compatible with current results from LHC Higgs searches. In the future, searches for heavy charged Higgs bosons in the channels $H^\pm\to\tau\nu$ and $H^\pm\to cs$ would be very interesting probes for such large couplings. 

%%%%%%%%%%%%%%%%%%%%%%%%%%%%%%%%%%%%%%%%%%%
%%%%%%%%%%%%%%%%%%%%%%%%%%%%%%%%%%%%%%%%%%%
\section{Conclusions}
\label{sec:conclusion}

In this paper we analysed the current phenomenological status of the charged Higgs boson in the Two Higgs Doublet Model. We considered constraints from Higgs searches at LEP and the LHC, measurements of the 125 GeV Higgs signal rates, and limits from flavour physics observables. 

Focussing first on the four 2HDM types with a $Z_2$-symmetric Yukawa structure, we showed that limits from direct charged Higgs searches at LEP and the LHC are very robust with respect to the variation of the 2HDM parameters, for each of the four types. The only exception to this observation is a possible mitigation of the exclusion from LHC charged Higgs searches --- mostly focussing on final states $\tau\nu$ and $tb$ from the decaying charged Higgs --- in the presence of a new, competing charged Higgs decay, namely either $H^\pm \to W^\pm A$ (with a light pseudoscalar Higgs $A$) and/or $H^\pm \to W^\pm h$ (with a light, non-standard Higgs boson $h$). We discussed in particular the latter decay mode, $H^\pm \to W^\pm h$, in the context of a 2HDM scenario with the SM-like, heavy $\CP$-even Higgs boson at $\sim 125\gev$, and showed that the branching fraction is generically sizeable if it is not kinematically suppressed. These observations thus warrant a dedicated LHC search program for the process $H^\pm \to W^\pm h$, with the charged Higgs boson $H^\pm$ either produced in top quark decays, or directly in association with a top- and bottom quark, and the neutral Higgs boson $h$ decaying into either $bb$ or $\tau\tau$. Such searches would give rise to important and complementary constraints on the charged Higgs sector of the 2HDM (and beyond).

In contrast, the constraints from neutral Higgs searches at the LHC strongly vary among the scenarios considered here, and thus do not  provide model-independent limits on the charged Higgs mass. Nevertheless, in specific scenarios, we find that neutral Higgs searches do give important constraints on the parameter space, and complement the constraints from direct charged Higgs searches.
Combining all the constraints from Higgs collider searches, we set a lower limit of $75\gev$ on the charged Higgs mass, independent of all other model parameters.

Flavour physics observables provide very strong constraints on the charged Higgs sector in the $Z_2$-symmetric scenarios, which depend (to very good approximation) only on $M_{H^+}$ and $\tb$. In particular, the inclusive decay $b\to s \gamma$ and, more generally, the $b\to s$ transitions lead to a robust exclusion of charged Higgs masses below $600\gev$ for types II and III, and of $650\gev$ for $\tan\beta < 1$ for all Yukawa types.

Yet, none of the standard 2HDM scenarios with a $Z_2$-symmetric Yukawa structure is able to explain the observed deviations in the decay $B\to D^{(*)} \ell \nu$ from the Standard Model. For this reason we analysed a scenario with generic charged Higgs couplings, and showed that, firstly, it is possible to fulfil the $B\to D^{(*)} \ell \nu$ constraint alone, leading to four distinct branches in the parameter space, and secondly, find parameter points that in addition also obey the $B\to\tau\nu$ and $D_s\to\tau\nu$ constraints (which depend on the same couplings). The acceptable scenarios have in general strongly enhanced couplings and a charged Higgs mass below $550\gev$, which may be probed at the LHC in searches for $H^\pm\to\tau\nu$ and $H^\pm\to cs$ in the future.

%%%%%%%%%%%%%%%%%%%%%%%%%%%%%%%%%%%%%%%%%%%
\section*{Acknowledgements}
TS thanks Philip Bechtle, Howie Haber, Mike Hance and Sven Heinemeyer for helpful discussions. The work of TS is partly funded by the US Department of Energy, grant number DE-SC0010107, and a Feodor-Lynen research fellowship sponsored by the Alexander von Humboldt foundation.

\bibliographystyle{h-physrev5}
\bibliography{chupdate-v2}

\end{document}